\documentclass[twocolumn,notitlepage,prb,superscriptaddress,longbibliography]{revtex4-2}
\usepackage{bbm}
\usepackage{amssymb}
\usepackage{algpseudocode}
\usepackage{algorithm}
\usepackage{amsmath}
\usepackage{braket}
\usepackage{booktabs}
\usepackage{color}
\usepackage{comment}
\usepackage{dsfont}
\usepackage{etoolbox}
\usepackage{epstopdf}
\usepackage[T1]{fontenc}
\usepackage{float}
\usepackage{graphicx}
\usepackage[colorlinks=true,linkcolor=blue,citecolor=blue,urlcolor=blue]{hyperref}
\usepackage{indentfirst}
\usepackage{latexsym,bm,euscript}
\usepackage{multirow}
\usepackage{natbib}
\usepackage{soul}
\usepackage{subfigure}
\usepackage{times}
\usepackage{titlesec}
\usepackage{tikz}
\usepackage{txfonts}
\usepackage{xspace}
\usepackage{xfrac}
\usepackage{etoolbox}
\usepackage{tikzMPS}
\usepackage{tikz-cd}
\usepackage{physics}
\usepackage{xfp}
\usepackage[normalem]{ulem}
\usepackage{tabularx}

\newcolumntype{Y}{>{\centering\arraybackslash}X}
\usepgflibrary {shadings}
\usetikzlibrary {arrows.meta}
\usetikzlibrary{shapes,positioning}

\tikzstyle{exec_red}=[rectangle, draw = red] 
\tikzstyle{exec}=[rectangle, draw]
\tikzstyle{decide_red}=[kite, kite vertex angles=150, draw = red]
\tikzstyle{decide}=[kite, kite vertex angles=150, draw]

\def\thetitle{Thermal Tensor Network Simulations of Lattice Fermions with Fixed Filling}

\definecolor{DarkBlue}{RGB}{50,120,180}
\definecolor{TextBlue}{RGB}{30,70,150}

\begin{document}
\title{\thetitle}

\author{Qiaoyi Li}
\affiliation{Institute of Theoretical 
Physics, Chinese Academy of Sciences, Beijing 100190, China}
\affiliation{School of Physical Sciences, University of Chinese Academy of Sciences, Beijing 100049, China}

\author{Dai-Wei Qu}
\affiliation{Arnold Sommerfeld Center for Theoretical Physics, Center for NanoScience, and Munich Center for Quantum Science and Technology, Ludwig-Maximilians-Universit\"at M\"unchen, 80333 Munich, Germany}
\affiliation{Institute of Theoretical 
Physics, Chinese Academy of Sciences, Beijing 100190, China}

\author{Bin-Bin Chen}
\affiliation{Peng Huanwu Collaborative Center for Research and Education, Beihang University, Beijing 100191, China}

\author{Tao Shi}
\email{tshi@itp.ac.cn}
\affiliation{Institute of Theoretical 
Physics, Chinese Academy of Sciences, Beijing 100190, China}

\author{Wei Li}
\email{w.li@itp.ac.cn}
\affiliation{Institute of 
Theoretical Physics, Chinese Academy of Sciences, Beijing 100190, China}
\affiliation{Hefei National Laboratory, University of Science and Technology of China, Hefei 230088, China}

\begin{abstract} 
Numerical simulations of strongly correlated fermions at finite temperature are essential for studying high-temperature superconductivity and other quantum many-body phenomena. The recently developed tangent-space tensor renormalization group (tanTRG) provides an efficient and accurate framework by representing thermal density operators as matrix product operators. However, the particle number generally varies during the cooling process. The conventional strategy of fine-tuning chemical potentials to reach a target filling is computationally demanding. Here we propose a fixed-$N$ tanTRG algorithm that stabilizes the average particle number by adaptively tuning the chemical potential within the imaginary-time evolution. We benchmark its accuracy on exactly solvable free fermions, and further apply it to the square-lattice Hubbard model. For hole-doped cases, we study the temperature evolution of charge and spin correlations, identifying several characteristic temperature scales for stripe formation. Our results establish fixed-$N$ tanTRG as an efficient and reliable tool for finite-temperature studies of correlated fermion systems.
\end{abstract}

\date{\today}

\maketitle

\section{Introduction}
The accurate simulation of strongly correlated electrons is a central challenge in modern computational condensed matter physics. Electron correlations give rise to a wealth of emergent phenomena in cuprates, such as high-$T_c$ superconductivity, pseudogap, and strange metal phases at finite temperature~\cite{Dagotto1994RMPCorrelated, Keimer2015Nquantum}. Correlated electrons also manifest in other prominent examples, such as quantum criticality in heavy fermion compounds~\cite{Si2010SHeavy, Coleman2015Heavy, Yang2016RoPiPTwo} and correlated topological phases in twisted materials~\cite{Andrei2020NMGraphene, Mak2022NNSemiconductor, Cai2023NSignatures, Park2023NObservation, Xu2023PRXObservation}. Gaining reliable insight into these complex quantum systems crucially relies on accurate many-body numerical simulations.

Simulating strongly correlated systems remains challenging, particularly at finite temperatures. Exact diagonalization is limited by the exponential growth of the Hilbert space and thus suffers from finite-size effects. Determinant quantum Monte Carlo (DQMC) methods~\cite{Scalapino1981PRBMonte, Blankenbecler1981PRDMonte}, though powerful, frequently encounter the sign problem, restricting their applicability at low temperatures and certain fillings~\cite{Iglovikov2015PRBGeometry}. Tensor network methods~\cite{Orus2014AoPpractical} provide a compelling alternative via low-entanglement approximations.
In particular, the density matrix renormalization group (DMRG)~\cite{White1992PRLDensity, Schollwoeck2005RMPdensity, Schollwoeck2011APDensity}, based on the matrix product states (MPS), has achieved remarkable success in ground-state studies.
A natural extension to finite temperatures leads to the matrix product operator (MPO) representation of thermal density operators~\cite{Verstraete2004PRLMatrix, Li2011PRLLinearized, Chen2018PRXExponential}. 
More recently, the tangent-space tensor renormalization group (tanTRG)~\cite{Li2023PRLTangent} has further advanced this framework by incorporating the time-dependent variational principles (TDVP)~\cite{Haegeman2011PRLTime, Haegeman2016PRBUnifying} into the imaginary-time evolution, enabling efficient and accurate simulations of quasi-one-dimensional systems~\cite{Wang2023PRLPlaquette, Qu2024PRLPhase, Qu2024PRLBilayer, Gao2024PRBDouble, Lv2025NCQuantum, Chen2026SBFractional}.

However, in contrast to ground-state DMRG, where the particle number can be fixed exactly by incorporating charge quantum numbers in the MPS, tanTRG---as well as other finite-temperature tensor network methods formulated in the grand canonical ensemble (GCE)---cannot directly control particle number during imaginary-time evolution.
While in most contexts, filling serves as a more intrinsic parameter for phase diagrams than the chemical potential.
For example, in the study of SC it is common to refer to the temperature-doping phase diagram~\cite{Qu2024PRLPhase,Qu2024PRLBilayer}, whereas in investigations of topological phases one often considers the temperature-filling phase diagram~\cite{Lu2024PRLThermodynamic,Chen2026SBFractional}.
To achieve a target particle number, dynamical tuning of the chemical potential has been proposed in QMC~\cite{Speidel2006JoCTaCAutomatic, Miles2022PREDynamical} and variational Gaussian states~\cite{Shi2020PRLVariational} methods, but it remains unexplored in the context of thermal tensor network simulations.
Traditionally, one begins with a trial chemical potential and then repeatedly adjusts it based on the resulting particle number after a full cooling process. Such a tedious manual search relies heavily on experience and adds substantial computational cost. Moreover, tracking the temperature evolution of a physical quantitiy at a fixed filling requires collecting data from multiple computations with different fixed chemical potentials, thereby multiplying the total cost by the number of independent computations. The difficulty becomes even more severe in compressible states, such as metals and superconductors, where the particle number is highly sensitive to the chemical potential.

To address this gap, we propose a fixed-$N$ tanTRG algorithm specially designed for charge-conserving systems. It adaptively tunes the chemical potential within the imaginary-time evolution framework, thereby controlling the particle number without the need for post-hoc chemical potential sweeps.
Our approach utilizes the geometric structure of the MPO manifold and incorporates a feedback mechanism based on the Riemannian gradient of the particle number operator, effectively stabilizing the system at the desired filling during the cooling process.
The remainder of this paper is organized as follows. In Sec.~II, we review the MPO-based finite-temperature formalism and discuss the geometric structure of tangent-space approaches. In Sec.~III, we introduce our adaptive chemical-potential tuning scheme and present the workflow of the fixed-$N$ tanTRG algorithm. In Sec.~IV, we first benchmark our method on exactly solvable non-interacting fermions to verify its accuracy, and then in Sec.~V apply it to the square-lattice Hubbard model to investigate the charge and spin stripes in the low-temperature regime.

\section{MPO-based finite temperature simulations}
\subsection{Bilayer MPO representation for thermal density operator}
A finite-temperature equilibrium state is completely characterized by the thermal density operator $e^{-\beta H}/Z$, where $H$ is the system Hamiltonian, $\beta = 1/T$ is the inverse temperature, and $Z = {\rm Tr}\left[e^{-\beta H}\right]$ is the partition function. 
Following purification via the Choi isomorphism~\cite{Choi1972CJoMPositive}, the expectation value of an observable $O$ can be expressed as
\begin{equation}
     \expectationvalue{O}_\beta = \frac{\expectationvalue{O}{\rho(\beta/2)}}{\braket{\rho(\beta/2)}},
\end{equation} 
where $\ket{\rho(\beta/2)} = e^{-\beta H/2}\ket{\rho_0}$ is the unnormalized purified density operator, obtained through imaginary-time evolution
\begin{equation}
     \frac{\mathrm{d}}{\mathrm{d}\tau}\ket{\rho(\tau)} = -H\ket{\rho(\tau)}
     \label{Eq:ImaginaryTimeEvolution}
\end{equation}
over a duration of $\beta/2$, starting from the purified identity operator $\ket{\rho_0}$ corresponding to infinite temperature.

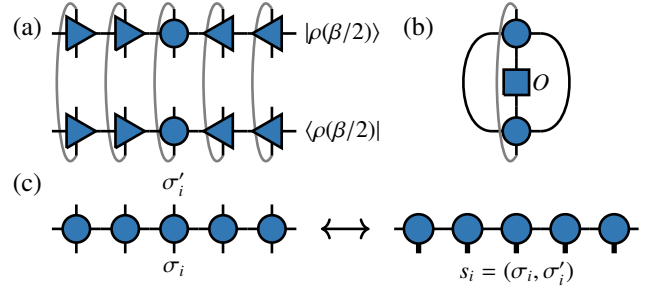
\begin{figure}[tbp]
     \centering
     \begin{tikzpicture}[scale = 0.65] 
         \node [below] at (0, -1.5) {\normalsize (a)};
         \draw [gray, line width = 1] (1, -1.5) [out = 120, in = 90] to (0.6, -3) [out = -90, in = -120] to (1, -4.5);
         \draw [gray, line width = 1] (2, -1.5) [out = 120, in = 90] to (1.6, -3) [out = -90, in = -120] to (2, -4.5);
         \draw [gray, line width = 1] (3, -1.5) [out = 120, in = 90] to (2.6, -3) [out = -90, in = -120] to (3, -4.5);
         \draw [gray, line width = 1] (4, -1.5) [out = 120, in = 90] to (3.6, -3) [out = -90, in = -120] to (4, -4.5);
         \draw [gray, line width = 1] (5, -1.5) [out = 120, in = 90] to (4.6, -3) [out = -90, in = -120] to (5, -4.5);
         \MPOTriaL{1}{-2}{}
         \MPOTriaL{1}{-4}{}
         \MPOTriaL{2}{-2}{}
         \MPOTriaL{2}{-4}{}
         \MPOCirc{3}{-2}{}
         \MPOCirc{3}{-4}{}
         \MPOTriaR{4}{-2}{}
         \MPOTriaR{4}{-4}{}
         \MPOTriaR{5}{-2}{}
         \MPOTriaR{5}{-4}{}
         \node [right] at (5.5, -2) {$\ket{\rho(\beta/2)}$};
         \node [right] at (5.5, -4) {$\bra{\rho(\beta/2)}$};
          
         \node [below] at (8, -1.5) {\normalsize (b)};
         \MPOCirc{10}{-2}{}
         \MPOCirc{10}{-4}{}
         \draw [gray, line width = 1] (10, -1.5) [out = 120, in = 90] to (9.6, -3) [out = -90, in = -120] to (10, -4.5);
         \draw [line width = 1] (9.5, -4) [out = 180, in = 180] to (9.5, -2);
         \draw [line width = 1] (10.5, -4) [out = 0, in = 0] to (10.5, -2);
         \draw [line width = 1] (10, -3.5) to (10, -2.5);
         \Squa{10}{-3}{}
         \node at (10.5, -3) {$O$};

          \node [above] at (0, -5.5) {\normalsize (c)};
          \MPOCirc{1}{-6}{}
          \MPOCirc{2}{-6}{}
          \MPOCirc{3}{-6}{}
          \node [below] at (3, -6.5) {$\sigma_i$};
          \node [above] at (3, -5.5) {$\sigma_i^\prime$};
          \MPOCirc{4}{-6}{}
          \MPOCirc{5}{-6}{}
          
          \draw [<->, line width = 1] (6, -6) -- (7, -6);
          
          \MPSCirc[2]{8}{-6}{}
          \MPSCirc[2]{9}{-6}{}
          \MPSCirc[2]{10}{-6}{}
          \node [below] at (10, -6.5) {$s_i = (\sigma_i,\sigma_i^\prime)$};
          \MPSCirc[2]{11}{-6}{}
          \MPSCirc[2]{12}{-6}{}
     \end{tikzpicture}
     \caption{(a) Bilayer MPO representation of the thermal density operator, which can be viewed as the reduced density operator of the purified state $\ket{\rho(\beta/2)}$ after tracing out the auxiliary indices. The triangles indicate the directions of canonical forms. (b) Tensor network used to compute the expectation value of an on-site observable located at the canonical center. (c) Illustration of the Choi isomorphism, which maps an MPO to an MPS with enlarged local Hilbert spaces.}
     \label{Fig2}
\end{figure}

We represent $\ket{\rho(\beta/2)}$ as a MPO[c.f. Fig.~\ref{Fig2}(a)] where the lower indices correspond to the physical spaces, and the upper indices represent the auxiliary spaces. The density operator on the enlarged Hilbert space is then expressed as a bilayer MPO~\cite{Dong2017PRBBilayer}. Tracing out the auxiliary indices produces the reduced density operator, which corresponds to the thermal density operator $e^{-\beta H}$.

An MPO can be viewed as an MPS under the Choi isomorphism, e.g., for the $i$'th local tensor, we map the two physical indices $\sigma_i$ and $\sigma_i'$ (living in the bra and ket spaces, $\langle\sigma_i'|$ and $|\sigma_i\rangle$) into a fused physical index $s_i=(\sigma_i,\sigma_i')$ (living in a ket space $|s_i\rangle=|\sigma_i\rangle\otimes|\sigma_i'\rangle$) 
[c.f. Fig.~\ref{Fig2}(c)]. 
This equivalence allows us to use it as a dictionary to translate MPS-based techniques into their MPO-based counterparts.
For example, we define an MPO local tensor as left(right)-canonical if and only if the corresponding MPS tensor is left(right)-canonical.
This formulation enables expectation value computations to follow the same procedures commonly used in ground-state DMRG calculations, as illustrated in Fig.~\ref{Fig2}(b).

\subsection{Tangent space projectors for the MPO manifold} 
\begin{figure}[tbp]
     \centering
     \begin{tikzpicture}[scale = 0.65] 
          \node [below] at (0, 1) {\normalsize (a)};
          \AMPSTriaL{1}{0}{}
          \MPSTriaL{1}{-1}{}
          \draw [line width = 1] (1.5, 0) -- (2, 0);
          \draw [line width = 1] (1.5, -1) -- (2, -1);
          \AMPSTriaL{2.5}{0}{}
          \MPSTriaL{2.5}{-1}{}
          \draw [line width = 1] (3, 0) -- (3.5, 0);
          \draw [line width = 1] (3, -1) -- (3.5, -1);
          \AMPSTriaL{4}{0}{}
          \BEnvR{4.5}{-1}{0}{}
          \MPSTriaL{4}{-1}{}
          \draw [line width = 2] (5.5, -1.5) -- (5.5, 0.5);
          \node [below] at (5.5, -1.5) {$i$};

          \AMPSTriaR{7}{0}{}
          \BEnvL{6.5}{-1}{0}{}
          \MPSTriaR{7}{-1}{}
          \draw [line width = 1] (7.5, 0) -- (8, 0);
          \draw [line width = 1] (7.5, -1) -- (8, -1);
          \AMPSTriaR{8.5}{0}{}
          \MPSTriaR{8.5}{-1}{}
          \draw [line width = 1] (9, 0) -- (9.5, 0);
          \draw [line width = 1] (9, -1) -- (9.5, -1);
          \AMPSTriaR{10}{0}{}
          \MPSTriaR{10}{-1}{}
          \draw [line width = 1] (10.5, 0) -- (11, 0);
          \draw [line width = 1] (10.5, -1) -- (11, -1);
          \AMPSTriaR{11.5}{0}{}
          \MPSTriaR{11.5}{-1}{}

          \node [below] at (0, -2) {\normalsize (b)};
          \AMPSTriaL{1}{-3}{}
          \MPSTriaL{1}{-4}{}
          \draw [line width = 1] (1.5, -3) -- (2, -3);
          \draw [line width = 1] (1.5, -4) -- (2, -4);
          \AMPSTriaL{2.5}{-3}{}
          \MPSTriaL{2.5}{-4}{}
          \draw [line width = 1] (3, -3) -- (3.5, -3);
          \draw [line width = 1] (3, -4) -- (3.5, -4);
          \AMPSTriaL{4}{-3}{}
          \MPSTriaL{4}{-4}{}
          \draw [line width = 1] (4.5, -3) -- (5.5, -3);
          \draw [line width = 1] (4.5, -4) -- (5.5, -4);
          
          \BEnvR{5.85}{-4}{-3}{}
          \draw [line width = 2] (5.5, -4.5) -- (5.5, -4);
          \node [below] at (5.5, -4.5) {$i$};
          \draw [line width = 2] (5.5, -3) -- (5.5, -2.5);
          \TriaL{5.5}{-3}{}
          \TriaL{5.5}{-4}{}

          \BEnvL{6.65}{-4}{-3}{}
          \draw [line width = 1] (7, -3) -- (8, -3);
          \draw [line width = 1] (7, -4) -- (8, -4);
          \draw [line width = 2] (7, -4.5) -- (7, -4);
          \node [below] at (7, -4.5) {$i+1$};
          \draw [line width = 2] (7, -3) -- (7, -2.5);
          \TriaR{7}{-3}{}
          \TriaR{7}{-4}{}

          \AMPSTriaR{8.5}{-3}{}
          \MPSTriaR{8.5}{-4}{}
          \draw [line width = 1] (9, -3) -- (9.5, -3);
          \draw [line width = 1] (9, -4) -- (9.5, -4);
          \AMPSTriaR{10}{-3}{}
          \MPSTriaR{10}{-4}{}
          \draw [line width = 1] (10.5, -3) -- (11, -3);
          \draw [line width = 1] (10.5, -4) -- (11, -4);
          \AMPSTriaR{11.5}{-3}{}
          \MPSTriaR{11.5}{-4}{}

          \node at (0, -5.7) {\normalsize (c)};
          \MPSTriaL{1}{-6.5}{}
          \MPOSqua{1}{-7.5}{}
          \AMPSTriaL{1}{-8.5}{}

          \draw [line width = 1] (1.5, -6.5) -- (1.6, -6.5);
          \draw [line width = 1] (1.5, -8.5) -- (1.6, -8.5);
          \MPSCirc{2.5}{-6.3}{}
          \MPOSqua{2.5}{-7.5}{}
          \draw [line width = 1] (1.5, -7.5) -- (2, -7.5);
          \node [above] at (2.5, -6) {$\ket{\rho_i^{\rm 1s}}$};
          \node [below] at (2.5, -8) {$H_i^{\rm 1s}$};

          \draw [line width = 1] (3.4, -6.5) -- (3.5, -6.5);
          \draw [line width = 1] (3.4, -8.5) -- (3.5, -8.5);
          \MPSTriaR{4}{-6.5}{}
          \MPOSqua{4}{-7.5}{}
          \draw [line width = 1] (3, -7.5) -- (3.5, -7.5);
          \AMPSTriaR{4}{-8.5}{}


          \node at (5.5, -5.7) {\normalsize (d)};
          \MPSTriaL{6.5}{-6.5}{}
          \MPOSqua{6.5}{-7.5}{}
          \AMPSTriaL{6.5}{-8.5}{}

          \draw [line width = 1] (7, -6.5) -- (7.5, -6.5);
          \draw [line width = 1] (7, -7.5) -- (7.5, -7.5);
          \draw [line width = 1] (7, -8.5) -- (7.5, -8.5);
          \MPSTriaL{8}{-6.5}{}
          \MPOSqua{8}{-7.5}{}
          \AMPSTriaL{8}{-8.5}{}

          \draw [line width = 1] (8.5, -6.5) -- (8.6, -6.5);
          \draw [line width = 1] (8.5, -7.5) -- (10.5, -7.5);
          \draw [line width = 1] (8.5, -8.5) -- (8.6, -8.5);

          \node [above] at (9.5, -6.2) {$\ket{\rho_i^{\rm b}}$};
          \node [below] at (9.5, -8) {$H_i^{\rm b}$};
          \draw [line width = 1] (9, -6.5) -- (10, -6.5);
          \Circ{9.5}{-6.5}{}

          \MPSTriaR{11}{-6.5}{}
          \MPOSqua{11}{-7.5}{}
          \AMPSTriaR{11}{-8.5}{}
          \draw [line width = 1] (10.4, -6.5) -- (10.5, -6.5);
          \draw [line width = 1] (10.4, -8.5) -- (10.5, -8.5);

     \end{tikzpicture}
     \caption{(a,b) Tensor network illustration of the tangent space projectors $P_i^{\rm 1s}$ and $P_i^{\rm b}$, respectively. The vertical line represents the identity operator acting on the $i$-th site. (c) Local renormalized state at the canonical center $\ket{\rho_i^{\rm 1s}}$, and the corresponding 1-site effective Hamiltonian $H_i^{\rm 1s}$ after the tangent space projection. (d) Similar to (c), but in the bond-canonical form.}
     \label{Fig3}
\end{figure}
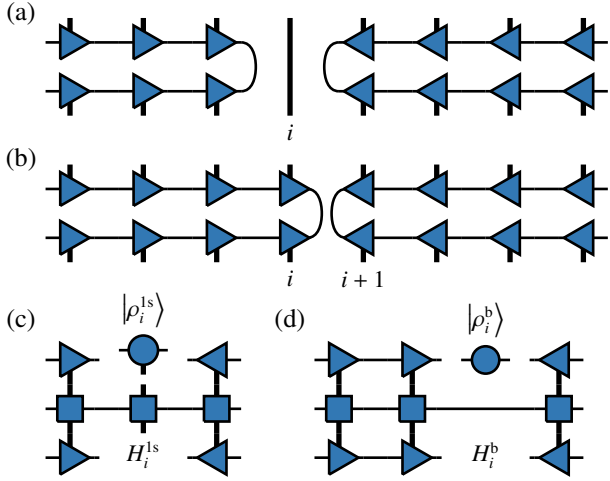

The geometric structure of the MPS manifold is well established~\cite{Haegeman2014JoMPGeometry,Gleis2022PRBProjector}. All the full-rank MPS's with a fixed bond dimension $D$ form a submanifold embedded in the Hilbert space.
Consequently, the tangent space $T_{\ket{\Psi}}\mathcal{M}$ at any state $\ket{\Psi}$ can be embedded into $T_{\ket{\Psi}}\mathcal{H}$ and further viewed as a subspace of the Hilbert space via the natural isomorphism $T_{\ket{\Psi}}\mathcal{H} \simeq \mathcal{H}$, where $\mathcal{M}$ denotes the MPS manifold and $\mathcal{H}$ denotes the Hilbert space.
Under this equivalence, the corresponding projector $P:\mathcal{H} \rightarrow T_{\ket{\Psi}}\mathcal{M}$ can be decomposed as 
\begin{equation}
     P = \sum_{i=1}^{N_s} P_i^{\rm 1s} - \sum_{i=1}^{N_s-1}P_i^{\rm b}
\end{equation}
in the mixed gauge, where $N_s$ is the number of lattice sites, as well as the length of MPS, $P_i^{\rm 1s}$ and $P_i^{\rm b}$ are the projectors onto the renormalized spaces at the $i$-th site and bond, respectively~\cite{Haegeman2016PRBUnifying}. For simplicity, we omit subscripts, but it is important to note that the projectors depend on the base point $\ket{\Psi}$. 
Due to the Choi isomorphism, these projectors have analogous forms for the MPO manifold, as illustrated in Fig.~\ref{Fig3}(a,b). 

Within the representation capacity of a MPS (MPO) manifold with a finite bond dimension,
the optimal continuous evolution is described by the tangent-space approximation of the evolution equation, known as TDVP equation. For instance, the imaginary-time evolution Eq.~(\ref{Eq:ImaginaryTimeEvolution}) is optimally approximated as
\begin{equation}
     \frac{\mathrm{d}}{\mathrm{d}\tau}\ket{\rho(\tau)} = -PH\ket{\rho(\tau)}.
     \label{Eq:TDVP}
\end{equation} 

\subsection{Geometric view of the imaginary-time evolution} 
As the energy expectation $E = \expectationvalue{H} = \expectationvalue{H}{\rho}/\braket{\rho}$ forms the Rayleigh quotient~\cite{Horn2012Matrix} of the Hamiltonian, the energy gradient in the purified Hilbert space is given by 
\begin{equation}
     \overline{\nabla} E = \frac{2(H - E)\ket{\rho}}{\braket{\rho}} \in T_{\ket{\rho}}\mathcal{H}.
\end{equation}
The gradient is inversely proportional to the norm $\|\rho\|$.
For convenience, we therefore define a norm-independent one as $\nabla E = \|\rho\|P\overline{\nabla} E \in T_{\ket{\rho}}\mathcal{M}$, which corresponds to the gradient projected to the unit sphere of the MPO manifold.
The embedding structure ensures that $\nabla E$ is indeed the intrinsic gradient on the MPO manifold with induced metric.
Therefore, the imaginary-time TDVP Eq.~(\ref{Eq:TDVP}) governs the evolution of the normalized density operator as
\begin{equation}
    \frac{\mathrm{d}}{\mathrm{d}\beta}\frac{\ket{\rho(\beta/2)}}{\|\rho(\beta/2)\|} = \frac{-(PH - \expval{PH})\ket{\rho(\beta/2)}}{2\|\rho(\beta/2)\|} = -\frac{1}{4}\nabla E,
\end{equation}
which implies that the geometric interpretation of the imaginary-time evolution is a continuous energy gradient descent on the MPO manifold.
Accordingly, the evolution of expectation value of an operator $O$ is characterized by 
\begin{equation}
    \frac{\mathrm{d}\expval{O}}{\mathrm{d}\beta} = -\frac{1}{4}\expval{\nabla E, \nabla O},
    \label{Eq:TDVP_O}
\end{equation}
where $\nabla O \equiv \nabla \expval{O}$ and $\expval{\cdot, \cdot}$ denotes the Riemannian metric on MPO manifold\cite{Lee2012GTiMIntroduction}. 

\section{Adaptively controlling particle number under GCE}
\subsection{Imaginary-time-dependent chemical potential}
A common approach to control the particle number is to apply a Legendre transformation, expressed as
\begin{equation}
     \widetilde{H} = H - \mu N,
\end{equation}
where $\mu$ is the chemical potential and $N$ is the total particle number operator.
This modification yields $e^{-\beta \widetilde{H}} = e^{-\beta (H - \mu N)} \equiv e^{-\alpha N -\beta H}$, indicating that the thermal density operator in GCE can be obtained using the standard purification approach with the modified Hamiltonian $H \rightarrow \widetilde{H}$. 

When the Hamiltonian conserves particle number, i.e. $[H, N] = 0$, we have 
\begin{equation}
     e^{-\alpha N - \beta H} = \exp\left[
          -\int_0^\beta \left[H - \mu(\tau)N\right]\mathrm{d}\tau
     \right]
\end{equation}
for any time-dependent $\mu(\tau)$, provided that
\begin{equation}
     \int_0^\beta \mu(\tau) \mathrm{d}\tau = -\alpha.
     \label{Eq:TimeDependentMu}
\end{equation}
This implies the system can be cooled along any path on the $\alpha-\beta$ plane without, in principle, altering the final result.
A similar property holds on the $\langle N\rangle-\beta$ plane, since $\langle N \rangle$ is uniquely determined by $\alpha$.
In particular, the constant-particle-number path naturally emerges as a viable option and is useful in practical studies\cite{Qu2024PRLPhase, Qu2024PRLBilayer,Chen2026SBFractional}.

\subsection{Dynamical selection of chemical potential}
Denoting the purified density operator as $\ket{\rho(\alpha/2, \beta/2)} \equiv e^{-\alpha N/2 -\beta H/2}\ket{\rho_0}$, the TDVP equation under GCE reads
\begin{equation}
     \frac{\mathrm{d}}{\mathrm{d}\beta}\frac{\ket{\rho(\alpha/2, \beta/2)}}{\|\rho(\alpha/2, \beta/2)\|} = -\frac{1}{4}\left(\nabla E - \mu \nabla N\right),
     \label{Eq:TDVPGCE}
\end{equation}
where $\nabla N$ is the norm-independent gradient of the particle number, analogous to $\nabla E$.
Correspondingly, the evolution of the average particle number is given by
\begin{equation}
     \left(\frac{\partial\expectationvalue{N}_\beta}{\partial\beta}\right)_\mu = -\frac{1}{4}\left(\langle \nabla N,\nabla E\rangle - \mu\langle \nabla N,\nabla N\rangle \right).
     \label{Eq:ParticleNumberFlow}
\end{equation}  
In principle, the particle number can be fixed by setting $(\partial\expectationvalue{N}_\beta/\partial\beta)_\mu = 0$~\cite{Shi2020PRLVariational}.
However, in practice, it will still drift due to a finite step length, as the derivative $(\partial\expectationvalue{N}_\beta/\partial\beta)_\mu$ varies during an integration step. To address this, we propose selecting $\mu$ according to
\begin{equation}
       \left(\frac{\partial\expectationvalue{N}_\beta}{\partial\beta}\right)_\mu = \frac{N_{\rm target} - \expectationvalue{N}_\beta}{\delta\beta},
\end{equation}
which leads to
\begin{equation}
     \mu = \frac{\delta\beta\langle \nabla N,\nabla E\rangle + 4N_{\rm target} - 4\expectationvalue{N}_\beta}{\delta\beta\langle \nabla N,\nabla N\rangle},
     \label{Eq:SelectMu}
\end{equation}
where $N_{\rm target}$ is the target particle number and $\delta\beta$ is the step length for the next TDVP sweep.
This approach implements a negative feedback mechanism, ensuring that the particle number is trapped around the target value.

To evaluate $\langle \nabla N,\nabla E\rangle$ within the MPO framework, we note that the Riemannian metric is given by the pullback of the inner product in the ambient Hilbert space. Consequently, substituting
$\nabla E = 2P(H - E)\ket{\rho}/\|\rho\|$ and $\nabla N = 2P(N - \expectationvalue{N})\ket{\rho}/\|\rho\|$ yields
\begin{equation}
     \langle \nabla N,\nabla E\rangle = 4\left({\rm Re}\expectationvalue{NPH} - \expectationvalue{N}\expectationvalue{H}\right).
\end{equation}
Using the decomposition of tangent-space projector, we have
\begin{equation}
     \begin{aligned}
          \expectationvalue{NPH} =& \sum_{i=1}^{N_s} \expectationvalue{NP_i^{\rm 1s}H} - \sum_{i=1}^{N_{s}-1} \expectationvalue{NP_i^{\rm b}H}\\
          =& \sum_{i=1}^{N_s} \frac{\expectationvalue{N_i^{\rm 1s}H_i^{\rm 1s}}{\rho_i^{\rm 1s}}}{\braket{\rho_i^{\rm 1s}}} - \sum_{i=1}^{N_{s}-1} \frac{\expectationvalue{N_i^{\rm b}H_i^{\rm b}}{\rho_i^{\rm b}}}{\braket{\rho_i^{\rm b}}},
     \end{aligned}
     \label{Eq:LocalConnection}
\end{equation}
where the second equality follows from the definitions of the states ($\ket{\rho_i^{\rm 1s}}$, $\ket{\rho_i^{\rm b}}$) and operators ($N_{i}^{\rm 1s}$, $N_{i}^{\rm b}$, $H_{i}^{\rm 1s}$, $H_{i}^{\rm b}$) in the local renormalized spaces, as illustrated in Fig.~\ref{Fig3}(c,d).

\subsection{Fixed-$N$ tanTRG algorithm}
By integrating the techniques described above with tanTRG, we propose a fixed-$N$ tanTRG algorithm for simulating charge-conserved fermion systems at finite temperature with a fixed target particle number, as illustrated in the flowchart in Fig.~\ref{Fig:Flowchart}.
We begin by initializing a high-temperature thermal density operator $e^{-\alpha_0 N/2 - \beta_0 H/2}$ with a small $\beta_0$ using the series-expansion thermal tensor network (SETTN) approach~\cite{Chen2017PRBSeries}, where the factor $1/2$ arises from the bilayer representation.
In the high-temperature limit, the system is only weakly influenced by the Hamiltonian, allowing the initial $\alpha_0$ to be estimated by locally solving
\begin{equation}
     \frac{{\rm Tr}\left[\hat{n} e^{-\alpha_0 \hat{n}}\right]}{{\rm Tr}\left[e^{-\alpha_0 \hat{n}}\right]} = n_{\rm target},
\end{equation}
where $\hat{n}$ is the local particle number operator and $n_{\rm target}$ is the target particle number per site.

We then proceed with the standard tanTRG algorithm to cool down the system, i.e., following the imaginary-time evolution equation of the unnormalized density operator, while adaptively using the chemical potential determined by Eq.~\eqref{Eq:SelectMu}.
In the initial stage, a 2-site algorithm or controlled bond expansion (CBE)~\cite{Gleis2023PRLControlled,Li2024PRLTime} is required to expand and further optimize the renormalized basis residing in the auxiliary bonds.
Moreover, a logarithmically scaled $\beta$ sequence is used, analogous to the exponential tensor renormalization group~\cite{Chen2018PRXExponential}, to rapidly reach low-temperature regime without introducing significant projection errors.
After that, we switch to a linearly scaled $\beta$ sequence to avoid uncontrolled Lie-Trotter errors due to indefinitely increasing step length in exponential cooling~\cite{Li2023PRLTangent}.

Before each TDVP sweep, a preliminary sweep is performed to determine the chemical potential $\mu$.
More specifically, we first apply the effective Hamiltonian and particle number operator to the local tensor at each site $i$, resulting $H_{i}^{\rm 1s}\ket{\rho_i^{\rm 1s}}$ and $N_{i}^{\rm 1s}\ket{\rho_i^{\rm 1s}}$, whose inner product renders the corresponding term in Eq.~\eqref{Eq:LocalConnection}. Similar procedure can be performed on each bond. 
By performing a left-to-right or right-to-left sweep to accumulate all these terms,
we obtain $\langle \nabla N,\nabla E\rangle$ and $\langle \nabla N,\nabla N\rangle$. 
As the expectation values $\expectationvalue{N}$ and $\expectationvalue{H}$ can be computed incidentally during this process, it is sufficient to determine $\mu$ using Eq.~\eqref{Eq:SelectMu}.
There is a computational overhead in fixed-$N$ scheme, about $1/(2K)$ of the baseline TDVP evolution. In a TDVP sweep, applying the effective Hamiltonian is performed $K-1$ times per site to obtain the $K$-dimensional Krylov space and constitutes the primary computational cost.

As mentioned earlier, the actual particle number after a TDVP sweep may still deviate from the target value.
When the deviation exceeds a preset tolerance, an additional evolution by $e^{-\delta \alpha N/2}$ is applied to restore the particle number to the target.
This adjustment is implemented using a Newton iteration method, which involves several time-evolving block decimation (TEBD)~\cite{Vidal2003PRLEfficient,Vidal2004PRLEfficient} sweeps.
The gradient used in the Newton iteration can be initially estimated as
\begin{equation}
     \left(\frac{\partial\expectationvalue{N}}{\partial\alpha}\right)_\beta = - \frac{1}{4}\langle \nabla N, \nabla N\rangle
\end{equation}
and subsequently refined using numerical differentiation after each iteration.
Note that no Trotter error is introduced in this step, as the particle number operators on different sites commute with each other.

\begin{figure}
     \centering 
     \begin{tikzpicture}[-stealth,scale = 1] 
          \node[exec] (SETTN) at (0, 0) {SETTN initialization};
          \node[exec_red] (estimate) [below=0.5 of SETTN] {estimate $\mu$ using Eq.~\eqref{Eq:SelectMu}};
          \node[exec] (TDVP) [below=0.5 of estimate] {imaginary-time TDVP sweep};
          \node[decide_red] (check) [below=0.5 of TDVP] {$|n - n_{\rm target}| < {\rm tol}?$};
          \node[exec_red] (TEBD) [right=0.5 of check] {TEBD correction};
          \node[exec] (Obs) [below=0.5 of check] {compute physical quantities};
          \node[decide] (beta) [below=0.5 of Obs] {$\beta \geq \beta_{\rm max}?$};
          \node[exec] (end) [below=0.5 of beta] {End};

          \draw (SETTN) -- (estimate);
          \draw (estimate) -- (TDVP);
          \draw (TDVP) -- (check);

          \draw (check) -- node[right] {Y} (Obs);
          \draw (check) -- node[above] {N} (TEBD);

          \draw (TEBD) |- (Obs);

          \draw (Obs) -- (beta);

          \draw (beta) -- node[right] {Y} (end);
          \draw (beta.west) -- node[above] {N} ++(-1.5,0) |-  (estimate);
     \end{tikzpicture}
     \caption{Flowchart of the fixed-$N$ tanTRG algorithm. Red blocks indicate the additional procedures necessary for fixing the   particle number.}
     \label{Fig:Flowchart}
\end{figure}

\subsection{Computation of thermodynamic quantities}
The internal energy can be directly extracted during the TDVP process as $E = \expectationvalue{H_{1}^{\rm 1s}}{\rho_1^{\rm 1s}}$ [c.f. Fig.~\ref{Fig3}(c)].
For other thermodynamic quantities, such as the free energy and entropy, we have
\begin{equation}
     F = -\frac{1}{\beta}\left(\ln \Xi + \alpha \langle N\rangle\right),
\end{equation} 
and 
\begin{equation}
     S = \beta\left(E - F\right),  
\end{equation} 
where $\Xi \equiv {\rm Tr}\left[e^{-\alpha N -\beta H}\right] = \innerproduct{\rho(\alpha/2,\beta/2)}{\rho(\alpha/2,\beta/2)}$ is the normalization factor that can be tracked during the time evolution.
It should be emphasized that all thermal quantities are defined in GCE and thus include contributions from particle number fluctuations, even though the average particle number $\langle N\rangle$ is fixed.  

The constant-particle-number specific heat is defined as
\begin{equation}
     C_N = (\partial E/\partial T)_{\langle N\rangle} = (\partial S/\partial \ln T)_{\langle N\rangle}.
\end{equation}
In practical calculations, numerically differentiating $(\partial S/\partial \ln T)_{\expval{N}}$ is more stable against numerical deviation from the target particle number at low temperatures. This can be understood from the estimates $(\partial E / \partial \expval{N})_{T} = \mu - T(\partial \mu/\partial T)_{\langle N\rangle}$ and $T(\partial S / \partial \expval{N})_T = -T(\partial \mu/\partial T)_{\langle N\rangle}$.

It is worth noting that $C_N$ differs from the constant-chemical-potential specific heat $C_\mu = (\partial (E - \mu\langle N\rangle)/\partial T)_\mu$ that is more commonly computed in previous fixed-$\mu$ grand-canonical algorithms~\cite{Chen2018PRXExponential, Li2023PRLTangent}.
Both $C_\mu$ and $C_N$ can be evaluated in tangent space, and the physical meanings of them are energy fluctuations with corrections due to particle number fluctuations, as detailed in Appendix~\ref{Appendix:SpecificHeat}. With the tangent-space formula, we obtain the relationship
\begin{equation}
     C_\mu= C_N + \frac{N_{\rm s} \chi_{\rm c}}{T}\left(\mu - \mu(\beta)\right)^2,
     \label{Eq:C_relation}
\end{equation}
where $\chi_{\rm c} = \left(\partial n/\partial \mu\right)_T$ is the charge susceptibility, $N_{\rm s}$ denotes the number of sites, and $\mu(\beta)$ represents the imaginary-time-dependent chemical potential in Eq.~\eqref{Eq:TimeDependentMu} that maintains a constant particle number during cooling. 

\section{Benchmark on non-interacting fermions}
We begin by benchmarking the fixed-$N$ tanTRG algorithm on a non-interacting spinless fermion chain, whose Hamiltonian reads
\begin{equation}
     H = -t\sum_{i} \left(c_i^\dagger c_{i+1}^{\phantom{\dagger}} + {\rm h.c.}\right),
\end{equation}
where $t=1$ is set as the energy unit. 
It should be noted that, although this model is exactly solvable, it constitutes a nontrivial technical benchmark for tensor network methods, as the metallic states possess high entanglement.

\begin{figure}[tbp]
     \centering
     \includegraphics[width = \linewidth]{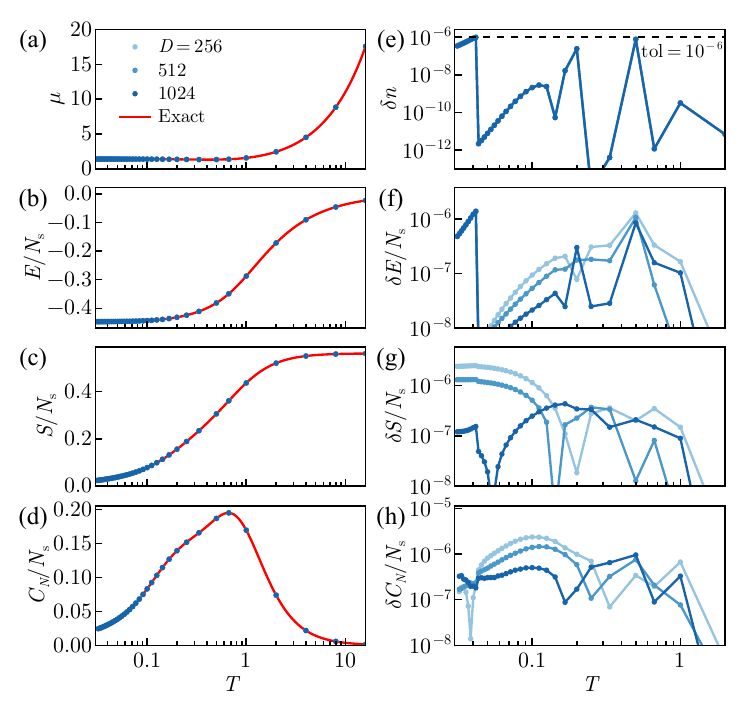}
     \caption{Benchmark of thermodynamic quantities for a spinless fermion chain of length 64 with open boundary conditions (OBC), where the filling is fixed at $n_{\rm target} = 3/4$ across all temperatures. (a-d) Temperature dependence of the chemical potential $\mu$, energy $E$, free energy $F$, entropy $S$, and specific heat $C_N$, respectively, normalized by system size $N_{\rm s} = 64$. (e) Deviation of the actual filling and the target value $n_{\rm target} = 3/4$. The gray dashed line indicates the tolerance $10^{-6}$ that triggers the TEBD correction. (f-h) Absolute errors of the thermodynamic quantities shown in panels (b-d), compared with the exact solution.}
     \label{Fig:Benchmark_FF}
\end{figure}

We first consider a target filling of $n_{\rm target} = 3/4$, as a typical case where a non-constant chemical potential is required to maintain fixed filling during cooling, as shown in Fig.~\ref{Fig:Benchmark_FF}(a).
At high temperatures, entropy dominates the free energy, so the system favors a highly disordered half-filled state. In this regime, a divergent chemical potential $\mu \sim -\alpha_0T$ (with finite $\alpha_0 = \ln\frac{1 - n_{\rm target}}{n_{\rm target}}$), is required to drive the system toward the target filling.
The thermodynamic quantities, including energy $E$, entropy $S$, and specific heat $C_N$, are shown in Fig.~\ref{Fig:Benchmark_FF}(b-d).
Here, the specific $C_N$ is computed in tangent space via Eq.~(\ref{Eq:C_N}) to avoid errors from numerical differentiation.
Across all temperatures, these thermodynamic quantities show excellent agreement with the exact solution, confirming the correctness and reliability of the fixed-$N$ tanTRG algorithm. 
The absolute errors are shown in Fig.~\ref{Fig:Benchmark_FF}(f-h), which are systematically improved as the bond dimension $D$ increases. For the largest bond dimension $D = 1024$, the dominant error becomes deviation from the target filling, which is controlled with the set tolerance $10^{-6}$, as shown in Fig.~\ref{Fig:Benchmark_FF}(e).

\begin{figure}[tbp]
     \centering
     \includegraphics[width = \linewidth]{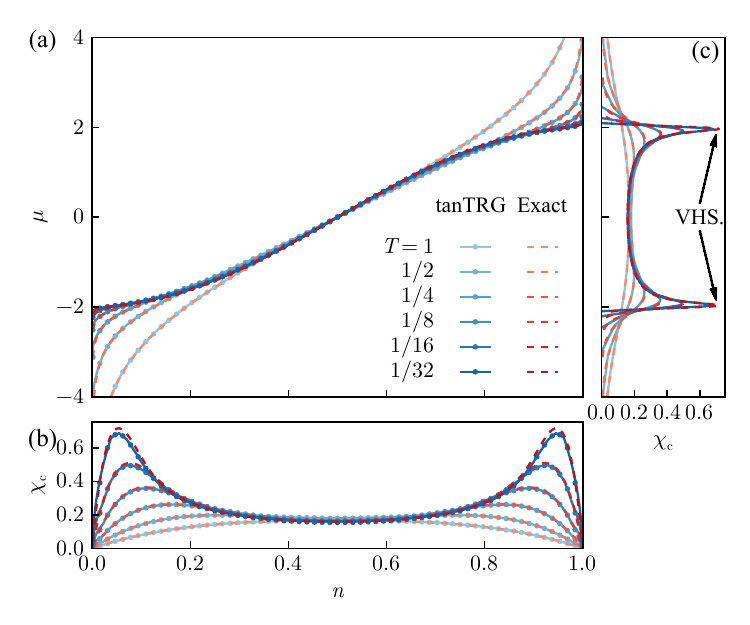}
     \caption{Benchmark of the EoS for an OBC spinless fermion chain of length 64, compared with the exact solution in TDL (red lines). (a) $n-\mu$ relation at several temperatures. Each marker corresponds to a target charge density obtained using the fixed-$N$ tanTRG algorithm. (b-c) Charge susceptibility $\chi_c$ as a function of $n$ and $\mu$, respectively. The lines are obtained via numerical differentiation, $\chi_c = (\partial n/\partial \mu)_T$, while the markers are obtained from charge fluctuations $\chi_c = \left(\expval{N^2} - \expval{N}^2\right) / (N_{\rm s}T)$. The bond dimension used is $D = 1024$, yielding fully converged results.}
     \label{Fig:EoS_FF}
\end{figure}

We next examine the equation of state (EoS)---the relationship among the charge density $n$, chemical potential $\mu$ and temperature $T$---which provides a basic characterization of a fermion system.
As shown in Fig.~\ref{Fig:EoS_FF}(a), using the fixed-N tanTRG algorithm, we obtain $n-\mu$ curves with $n$ uniformly scaled at all temperatures, which agree well with the exact solution $n(\mu, T) = \int_{-\pi}^\pi \frac{dk}{2\pi} \frac{1}{e^{(\epsilon_k-\mu)/T} + 1}$ in thermodynamic limit (TDL), where $\epsilon_k = -2\cos k$ is the dispersion.
We then compute the charge susceptibility $\chi_c$ and show its dependence on $n$ and $\mu$ in Fig.~\ref{Fig:EoS_FF}(b) and (c), respectively. When the chemical potential lies within the band, i.e. $-2 < \mu < 2$, $\chi_c$ decreases monotonically with temperature, exhibiting metallic behavior. In contrast, when $\mu$ lies outside the band, $\chi_c$ is strongly suppressed as the system is cooled, consistent with a band-insulator character.
For non-interacting fermions, $\chi_c$ approaches the density of states in the low-temperature limit and thus exhibits van Hove singularities (VHS) at the band edges ($\mu = \pm 2$).  

\section{Application: Square-lattice Hubbard model}
\subsection{Thermodynamic Quantities}
We next apply the fixed-$N$ tanTRG algorithm to the square-lattice Hubbard model~\cite{Hubbard1963PRSAElectron}, defined by
\begin{equation}
     H = -t\sum_{\langle i,j\rangle\sigma} \left(c_{i\sigma}^\dagger c_{j\sigma}^{\phantom{\dagger}} + \textrm{h.c.}\right) + U\sum_i n_{i\uparrow} n_{i\downarrow},
\end{equation}
where $\langle i,j\rangle$ denotes nearest neighbor (NN) sites.
Despite its simple form, Hubbard model involves rich physics and is widely studied as a minimal model for strongly correlated electrons~\cite{Arovas2022ARoCMPHubbard, Qin2022ARoCMPHubbard}.

\begin{figure}[tbp]
     \centering
     \includegraphics[width = \linewidth]{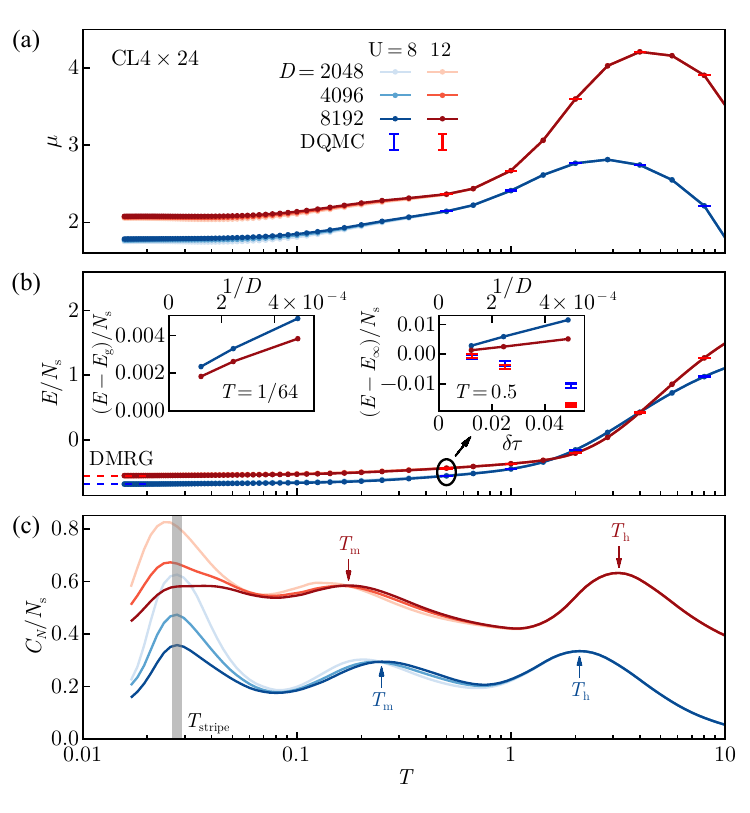}
     \caption{Thermodynamic quantities of the $\delta = 1/12$ hole-doped Hubbard model on 4$\times$24 cylinder, with typical parameters $U = 8$ and $U = 12$. (a) Chemical potential $\mu$ required to maintain the target doping, obtained from fixed-$N$ tanTRG and DQMC. (b) Per-site energy $E/N_{\rm s}$. Ground state energies $E_g/N_{\textrm{s}}$ from DMRG are indicated by horizontal dashed lines. The inset shows the convergence with bond dimension $D$ at $T =1/64$ (left) and $T = 0.5$ (right), where the offset $E_\infty$ is estimated by linearly extrapolating $1/D \rightarrow 0$.
     The Trotter step length in DQMC is set as $\delta\tau = 0.0125$ in main panel, whose effect is also examined at $T = 0.5$.
     (c) Specific heat $C_N/N_{\rm s}$ evaluated from numerical differentiation $(\partial S/\partial \ln T)_{\expval{N}}$. The data for $U=12$ are vertically offset by 0.3 for clarity. The arrows indicate the characteristic temperature scales $T_{\rm h}$ and $T_{\rm m}$, whereas the shaded band marks $T_{\rm stripe}$.}
     \label{Fig:Benchmark_Hubbard}
\end{figure}

Fig.~\ref{Fig:Benchmark_Hubbard}(a) shows the temperature dependence of the chemical potential $\mu$ that maintains the target doping $\delta = 1/12$. Adding a particle to a Hubbard system requires overcoming the on-site repulsion $U$, which leads to the peak feature in $\mu-T$ curves with peak value increasing with $U$. 
Note $(\partial n/\partial T)_\mu = - \chi_{\rm c}(\partial\mu/\partial T)_{\expval{N}}$, the positive slope below the peak yields $(\partial n/\partial T)_\mu < 0$, which means charged excitations around fermi level are dominated by holes, rather than electrons, consistent with the hole carrier fact in a slightly hole-doped Mott insulator.

At high temperatures ($T \geq 0.5$), where the sign problem remains moderate,
we benchmark our results against DQMC simulations employing dynamical tuning of the chemical potential~\cite{Miles2022PREDynamical}.
The per-site energy $E/N_{\rm s}$ is displayed in Fig.~\ref{Fig:Benchmark_Hubbard}(b), exhibiting excellent agreement with DQMC results at high temperatures, and approaching the DMRG ground-state energy smoothly upon cooling.
The data convergence with respect to bond dimension $D$ is shown in the inset panels of Fig.~\ref{Fig:Benchmark_Hubbard}(b). 
At $T = 0.5$, the per-site energy is overestimated by tanTRG due to finite bond dimension $D$, while DQMC underestimates it because of a finite Trotter step length $\delta\tau$.
We also estimate the accuracy at lowest simulated temperature $T = 1/64$ by comparing with DMRG ground-state energy $E_\textrm{g}/N_\textrm{s} = -0.6811$ ($U = 8$) and $-0.5504$ ($U = 12$), obtained with bond dimension $D = 16384$.
At both temperatures, fixed-$N$ tanTRG achieves an energy accuracy of order $10^{-3}$ for $D = 8192$ and can be systematically improved by increasing $D$. 

For a strongly correlated system like doped Hubbard model, the finite bond dimension $D$ becomes a major bottleneck in evaluating $C_N$.
This limitation is further exacerbated when using the tangent-space expression Eq.~\eqref{Eq:C_N}, as the associated energy fluctuations involve long-range correlations that are challenging to capture accurately within MPO representation.
Therefore, we first apply spline interpolation to the $S-\ln T$ curve and then evaluate $C_N = (\partial S/\partial \ln T)_{\expval{N}}$ using a centered finite difference, with results shown in Fig.~\ref{Fig:Benchmark_Hubbard}(c).
The specific heat exhibits a three-peak structure, with peak positions corresponding to different temperature scales, which will be discussed in Sec.~\ref{Sec:TemperatureScales}.

\subsection{Charge and Spin Stripes}
\begin{figure}[tbp]
     \centering
     \includegraphics[width = \linewidth]{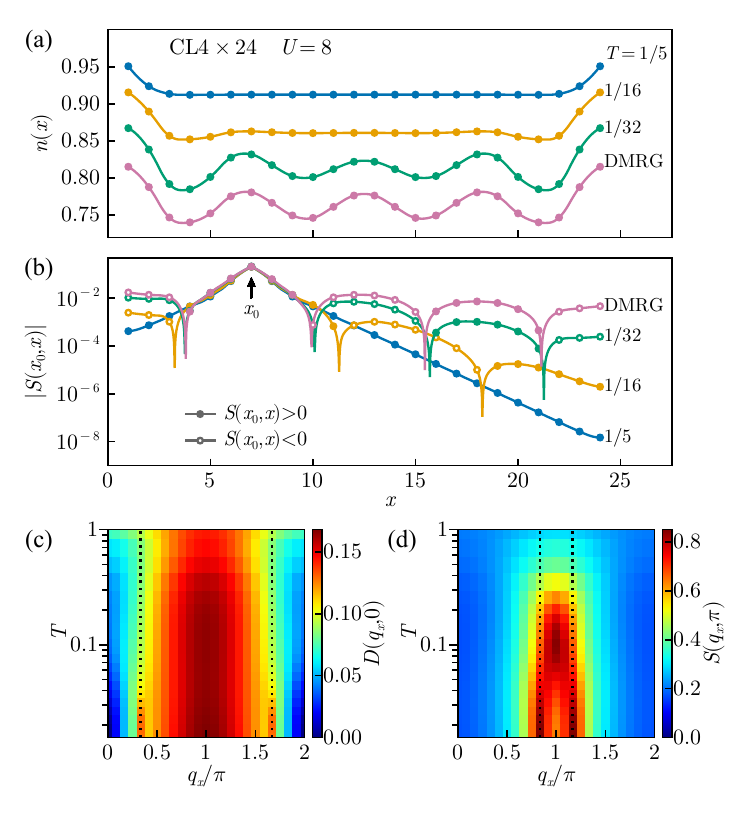}
     \caption{Charge and spin stripes in the $\delta = 1/12$ hole-doped Hubbard model on CL4$\times$24 cylinder with $U = 8$.
     (a) Charge density distribution $n(x)$ and (b) staggered spin correlations $S(x_0 = 7, x)$ at several representative temperatures. Each curve in (a) is vertically offset by $0.05$ successively for clarity.
     (c, d) Contour plot of the charge structure factor $D(q_x, 0)$ and spin structure factor $S(q_x, \pi)$ as functions of temperature $T$ and momentum $q_x$.
     Data in (a,b) are spline-interpolated to highlight the stripe periodicity.
     Vertical dotted lines in (c,d) indicate the characteristic wave vectors $q_x = 2\pi\delta/f$ (CDW) and $q_x = \pi(1-\delta/f)$ (SDW), with the filling fraction $f = 1/2$. All finite-temperature results are obtained with bond dimension $D = 8192$.}
     \label{Fig:Hubbard_Stripe}
\end{figure}

We next examine the temperature evolution of stripes in the Hubbard model.
Fig.~\ref{Fig:Hubbard_Stripe}(a,b) display the real-space charge density distribution $n(x) = \frac{1}{W}\sum_{y = 1}^W n(x, y)$ and staggered spin correlations $S(x_0, x) = \frac{1}{3W}\sum_{y = 1}^W (-1)^{|x-x_0|}\langle \mathbf{S}_{(x_0,y)}\cdot \mathbf{S}_{(x,y)}\rangle$, respectively, at several representative temperatures.
At high temperatures (e.g. $T = 1/5$), the system remains charge uniform and exhibits only short-range AFM correlations, except for a slightly enhanced density near the open boundaries.
Upon cooling, the boundary-induced charge oscillations propagate inward, and gradually evolve into a charge density wave (CDW) with a wavelength of $\lambda_{\rm CDW} = 6$.
This corresponds to the half-filled ($f = 1/2$) stripe reported in previous studies~\cite{Jiang2020PRRGround, Wietek2021PRXStripes}, where the filling fraction is defined as $f = \delta \lambda_\textrm{CDW}$~\cite{Xu2024SCoexistence}.
With the emergence of CDW, a $\pi$-phase shift develops in the spin correlations and eventually locks to the charge-density minima, consistent with the stripe picture.

This behavior is also reflected in the enhancement of the charge and spin structure factor
\begin{equation}
     D(\mathbf{q}) = \frac{1}{N_{\rm s}}\sum_{ij}e^{-i\mathbf{q}\cdot(\mathbf{r}_i - \mathbf{r}_j)}\expval{(n_i - n)(n_j - n)}
\end{equation}
and
\begin{equation}
     S(\mathbf{q}) = \frac{1}{3N_{\rm s}}\sum_{ij}e^{-i\mathbf{q}\cdot(\mathbf{r}_i - \mathbf{r}_j)}\expval{\mathbf{S}_i \cdot \mathbf{S}_j},
\end{equation}
at the CDW wave vector $\mathbf{q}_{\rm CDW} = (2\pi\delta/f, 0)$ and corresponding spin density wave (SDW) wave vector $\mathbf{q}_{\rm SDW} = (\pi - \pi\delta/f, \pi)$, respectively, as shown in Fig.~\ref{Fig:Hubbard_Stripe}(c, d). 
In addition to the low-temperature signatures of the half-filled stripe, a broad feature near $\mathbf{q} = (\pi, 0)$ persists in $D(\mathbf{q})$ over a wide temperature range, which is attributed to the high-energy charge excitations~\cite{Mai2023NCRobust}.

\subsection{Characteristic Temperatures Scales{\label{Sec:TemperatureScales}}}
The rich temperature dependence of spin and charge correlations gives rise to multiple characteristic temperature scales, summarized in Table~\ref{Tab:T}.
Upon cooling from the high-temperature limit, the first scale is the highest-temperature specific-heat peak, $T_{\rm h}$, which is associated with the suppression of double occupancy~\cite{Li2023PRLTangent} and is therefore approximately proportional to $U$.
Upon further cooling, short-range AFM correlations develop as a result of the effective spin-exchange interaction $J = 4t^2/U$, giving rise to a magnetic temperature scale $T_{\rm m}$, which is identified from the second specific heat peak.

\begin{table}[htbp]
     \centering
     \caption{Characteristic temperature scales of hole-doped Hubbard model on CL4$\times$24 cylinder. Numbers in parentheses indicate the estimated uncertainties, based on the differences between results with bond dimension $D = 4096$ and $D = 8192$, except for $T_{\rm h}$, which is converged with respect to $D$, while its accuracy is limited by interpolation from discrete temperature points.}
     
     \begin{tabularx}{\linewidth}{|Y|Y|Y|Y|Y|Y|}
          \hline 
          \rule{0em}{1.05em} $U$ & $\delta$ & $T_{\rm stripe}$ & $T_{\rm SDW}^\ast$ & $T_{\rm m}$ & $T_{\rm h}$\\
          \hline
          \rule{0em}{1.1em} 8 & 1/12 & 0.028(1) & 0.103(3) & 0.25(2) & 2.1\\
          \hline
          \rule{0em}{1.1em} 12 & 1/12 & 0.027(1) & 0.083(3) & 0.17(1) & 3.2\\  
          \hline
     \end{tabularx}
     \label{Tab:T}
\end{table}

\begin{figure}[tbp]
     \centering
     \includegraphics[width=\linewidth]{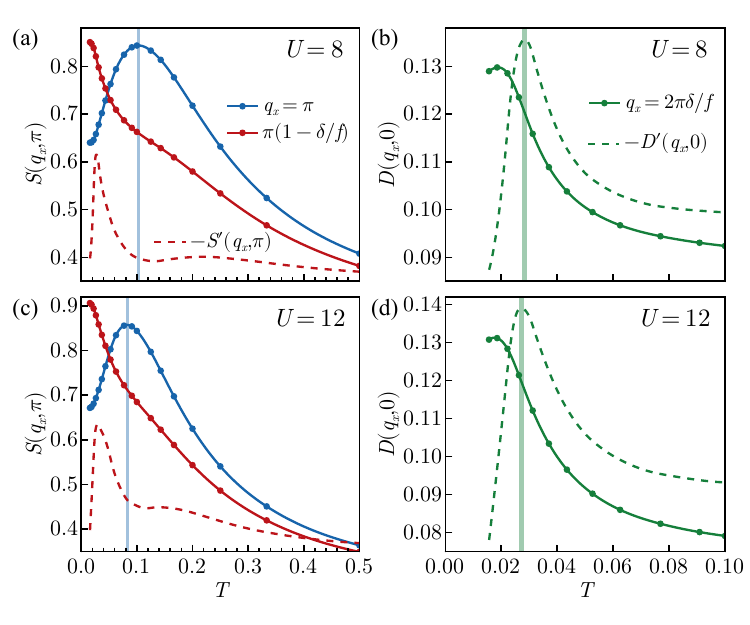}
     \caption{Temperature dependence of (a,c) spin structure factor $S(q_x, \pi)$ and (b,d) charge structure factor $D(q_x, 0)$ at relevant wave vectors of the Hubbard model on a CL$4\times24$ cylinder. The parameters are $U = 8$ for (a,b) and $U = 12$ for (c,d), with doping level $\delta = 1/12$ and bond dimension $D = 8192$.
     The dashed curves represent the temperature derivatives {$S^\prime(\mathbf{q}) = \partial S(\mathbf{q}) / \partial T$ and $D^\prime(\mathbf{q}) = \partial D(\mathbf{q}) / \partial T$ at relevant wave vectors}, respectively, in arbitrary units.
     Vertical shaded bands indicate the characteristic temperatures $T_{\rm SDW}^\ast$ (blue) and $T_{\rm stripe}$ (green), respectively.}
     \label{Fig:Hubbard_T}
\end{figure}

The establishment of AFM correlations enhances $S(\pi, \pi)$ upon cooling, until the emergence of SDW shifts the spin spectral weight away from $\mathbf{q}_{\rm AFM} = (\pi, \pi)$.
We therefore define $T_{\rm SDW}^\ast$, the onset temperature of SDW, as the temperature corresponding to the peak of $S(\pi, \pi)$, below which $S(\mathbf{q}_{\rm SDW})$ begins to increase rapidly, as shown in Fig.~\ref{Fig:Hubbard_T}(a,c).
Finally, the characteristic temperature of stripe, $T_{\rm stripe}$, is determined by the temperature at which $D(\mathbf{q}_{\rm CDW})$ increases most rapidly, i.e. the peak of $-\partial D(\mathbf{q}_{\rm CDW})/\partial T$, as shown in Fig.~\ref{Fig:Hubbard_T}(b,d). 
Note that a $\pi$-phase shift in spin correlations has already appeared between $T_{\rm stripe}$ and $T_{\rm SDW}^\ast$, but the SDW and CDW patterns remain unlocked, as exemplified by the case of $T = 1/16$ in Fig.~\ref{Fig:Hubbard_Stripe}(a, b).
While $T_m$ exhibits a clear $1/U$ dependence, both $T_{\rm SDW}^\ast$ and $T_{\rm stripe}$ show much weaker sensitivity to $U$. This behavior can be understood as a result of the stripe formation being governed by a balance between the kinetic energy ($\propto t$) and the magnetic energy ($\propto t^2/U$), and therefore not following a simple scaling in either $t$ or $U$ alone.

\section{Discussions}
In this work, we develop a fixed-$N$ tanTRG method that enables finite-temperature simulations of charge-conserving fermion systems in the grand canonical ensemble with controlled average particle number. 
By incorporating an adaptive feedback scheme for tuning the chemical potential directly into the imaginary-time evolution, our method achieves accurate stabilization of the filling.
Moreover, since the main additional computation involves applying the effective Hamiltonians to the renormalized states [c.f. Fig.~\ref{Fig3}(c)], this modification can be implemented straightforwardly within the original tanTRG framework. 
Compared with a single fixed-$\mu$ simulation, the additional computational cost is estimated to be $1/(2K)$ of the original one, where $K$ denotes the Krylov space dimension used in TDVP. Given that $K$ is typically of order 10, this represents only a moderate overhead.
In contrast, traditional fixed-$\mu$ method requires scanning over multiple $\mu$ values to locate the desired filling, making it significantly less efficient for simulations targeting a fixed particle number.

We benchmarked the method against exactly solvable non-interacting fermions, where it accurately reproduces thermodynamic quantities, the EoS and charge susceptibility. Applying the approach to the square-lattice Hubbard model, we further demonstrated its ability to capture subtle emergent many-electron states and phenomena, including temperature evolution of stripes at slight hole doping. These results establish fixed-$N$ tanTRG as a powerful and reliable algorithm for exploring finite-temperature physics of strongly correlated systems.

Beyond particle number in fermion systems, the strategy developed here can be generalized to other conserved quantities and their conjugate fields. Moreover, the fixed-$N$ scheme is not limited to tanTRG: it can be readily incorporated into other thermal tensor network methods based on imaginary-time evolution, such as finite-temperature projected entangled-pairs operators~\cite{Li2011PRLLinearized, Czarnik2012PRBProjected, Czarnik2016PRBVariational, Sinha2022PRBFinite, Zhang2025Finite}, as long as the gradient of the relevant observables, e.g., energy and particle number, can be estimated.

\textit{Acknowledgments}.--- Q.L. and W.L. are indebted to Jialin Chen for stimulating discussions. The tanTRG and DMRG computations are based on the open-source package FiniteMPS.jl~\cite{FiniteMPS.jl}, whose tensor backend is TensorKit.jl~\cite{Devos2025TensorKit.jl}. The DQMC simulations are performed using the open-source package SmoQyDQMC.jl~\cite{CohenStead2024SPCSmoQyDQMC.jl}. The source code and data that support this work are available in the GitHub repository Fixed-N-tanTRG~\cite{data_repository}.  
This work is supported by the National Natural Science Foundation of China (Grant Nos.~12534009, and 12447101), and the Innovation Program for Quantum Science and Technology (Grand No. 2021ZD0301900). We thank HPC at ITP-CAS for the technical 
support and generous allocation of CPU time.

\appendix

\section{Appendix: Tangent-space formulas for specific heats\label{Appendix:SpecificHeat}}
The constant-chemical-potential specific heat $C_\mu$ and constant-particle-number specific heat $C_N$ can be expressed in terms of the Riemannian metric of the gradient fields $\nabla E$ and $\nabla N$, as derived below.

As shown in Eq.~\eqref{Eq:TDVPGCE}, the imaginary-time evolution is governed by the gradient field $-\frac{1}{4}(\nabla E - \mu \nabla N)$.
Similar to Eq.~\eqref{Eq:TDVP_O}, for a fixed chemical potential $\mu$, the differential of an observable $O$ with respect to inverse temperature $\beta$ reads
\begin{equation}
     \left(\frac{\partial O}{\partial \beta}\right)_\mu = -\frac{\langle \nabla E, \nabla O\rangle - \mu \langle \nabla N, \nabla O\rangle}{4}.
\end{equation} 
Substituting $O = E - \mu \langle N\rangle$ yields
\begin{equation}
     \begin{aligned}
          C_\mu  \equiv& \left(\frac{\partial E - \mu \expval{N}}{\partial T}\right)_\mu =  -\frac{1}{T^2}\left(\frac{\partial E - \mu \expval{N}}{\partial \beta}\right)_\mu\\
          =& \frac{1}{4T^2}\left[
               \langle \nabla E, \nabla E - \mu \nabla N\rangle - \mu \langle \nabla N, \nabla E - \mu \nabla N\rangle
               \right]\\
          =& \frac{1}{4T^2}\left[\langle\nabla E, \nabla E\rangle - 2\mu \langle\nabla N, \nabla E\rangle + \mu^2\langle\nabla N, \nabla N\rangle\right].
     \end{aligned}
     \label{Eq:C_mu}
\end{equation}

To obtain $C_N$, we first determine the chemical potential $\mu$ that enforces a fixed particle number $\langle N\rangle$ according to
\begin{equation}
     \left(\frac{\partial \expval{N}}{\partial \beta}\right)_\mu = -\frac{\langle \nabla E, \nabla N\rangle - \mu \langle \nabla N, \nabla N\rangle}{4} = 0,
\end{equation}
which results in $\mu = \langle \nabla E, \nabla N\rangle/\langle \nabla N, \nabla N\rangle$.
Thus,
\begin{equation}
     \begin{aligned}
          C_N \equiv& \left(\frac{\partial E}{\partial T}\right)_{\expval{N}} = -\frac{1}{T^2}\left(\frac{\partial E}{\partial \beta}\right)_{\expval{N}}\\
          =& \frac{1}{4T^2}\left(\langle\nabla E, \nabla E\rangle - \mu \langle\nabla N, \nabla E\rangle\right)\\
          =& \frac{1}{4T^2}\left[\langle\nabla E, \nabla E\rangle - \frac{\langle\nabla N, \nabla E\rangle^2}{\langle\nabla N, \nabla N\rangle}\right].
     \end{aligned}
     \label{Eq:C_N}
\end{equation}
Note $\langle \nabla E, \nabla E\rangle = 4\left(\expval{HPH} - \expval{H}^2\right)$, both specific heat measure the energy fluctuations but with different corrections from particle number fluctuations due to the different derivative directions. 
Combining Eq.~\eqref{Eq:C_mu} and Eq.~\eqref{Eq:C_N}, we have 
\begin{equation}
     C_\mu = C_N + \frac{1}{4T^2}\langle \nabla N,\nabla N\rangle \left(\mu -  \frac{\langle\nabla N, \nabla E\rangle}{\langle\nabla N, \nabla N\rangle}\right)^2.
\end{equation}  
Note that the charge susceptibility is given by $\chi_c = \frac{\langle \nabla N, \nabla N\rangle}{4 N_s T}$, and $\frac{\langle \nabla N, \nabla E\rangle}{\langle \nabla N, \nabla N\rangle}$ is exactly the imaginary-time-dependent chemical potential $\mu(\beta)$ that maintains a constant particle number during cooling [see Eq.~\eqref{Eq:ParticleNumberFlow}], which leads to Eq.~\eqref{Eq:C_relation} in the main text.

\bibliography{./tanTRG_fixN.bib}

@misc{data_repository,
    title  = {{Fixed-N-tanTRG}},
    author = {Qiaoyi Li},
    year = {2026},
    publisher = {GitHub},
    howpublished = {\url{https://github.com/Qiaoyi-Li/Fixed-N-tanTRG.git}}
}

@misc{FiniteMPS.jl,
  title = {{FiniteMPS.jl}},
  author = {Qiaoyi Li},
  year = {2026},
  publisher = {GitHub},
  howpublished = {\url{https://github.com/Qiaoyi-Li/FiniteMPS.jl}} 
}

@Article{Shi2020PRLVariational,
  author    = {Shi, Tao and Demler, Eugene and Cirac, J. Ignacio},
  journal   = {Phys. Rev. Lett.},
  title     = {Variational Approach for Many-Body Systems at Finite Temperature},
  year      = {2020},
  month     = {Oct},
  pages     = {180602},
  volume    = {125},
  doi       = {10.1103/PhysRevLett.125.180602},
  issue     = {18},
  numpages  = {5},
  publisher = {American Physical Society},
  url       = {https://link.aps.org/doi/10.1103/PhysRevLett.125.180602},
}

@Article{Haegeman2016PRBUnifying,
  author    = {Jutho Haegeman and Christian Lubich and Ivan Oseledets and Bart Vandereycken and Frank Verstraete},
  journal   = {Phys. Rev. B},
  title     = {Unifying time evolution and optimization with matrix product states},
  year      = {2016},
  month     = {oct},
  number    = {16},
  pages     = {165116},
  volume    = {94},
  doi       = {10.1103/physrevb.94.165116},
  fjournal  = {Physical Review B},
  publisher = {American Physical Society ({APS})},
  ranking   = {rank5},
}

@Article{Haegeman2011PRLTime,
  author    = {Jutho Haegeman and J. Ignacio Cirac and Tobias J. Osborne and Iztok Pi{\v{z}}orn and Henri Verschelde and Frank Verstraete},
  journal   = {Phys. Rev. Lett.},
  title     = {{Time-Dependent Variational Principle for Quantum Lattices}},
  year      = {2011},
  month     = {aug},
  number    = {7},
  pages     = {070601},
  volume    = {107},
  doi       = {10.1103/physrevlett.107.070601},
  fjournal  = {Physical Review Letters},
  publisher = {American Physical Society ({APS})},
  ranking   = {rank4},
}

@Article{Haegeman2014JoMPGeometry,
  author    = {Jutho Haegeman and Micha\"el Mari\"en and Tobias J. Osborne and Frank Verstraete},
  journal   = {J. Math. Phys.},
  title     = {Geometry of matrix product states: Metric, parallel transport, and curvature},
  year      = {2014},
  month     = {feb},
  number    = {2},
  pages     = {021902},
  volume    = {55},
  doi       = {10.1063/1.4862851},
  fjournal  = {Journal of Mathematical Physics},
  publisher = {{AIP} Publishing},
  ranking   = {rank5},
}

@Article{Chen2018PRXExponential,
  author    = {Bin-Bin Chen and Lei Chen and Ziyu Chen and Wei Li and Andreas Weichselbaum},
  journal   = {Phys. Rev. X},
  title     = {{Exponential Thermal Tensor Network Approach for Quantum Lattice Models}},
  year      = {2018},
  month     = {sep},
  number    = {3},
  pages     = {031082},
  volume    = {8},
  doi       = {10.1103/physrevx.8.031082},
  fjournal  = {Physical Review X},
  publisher = {American Physical Society ({APS})},
}

@Article{Xu2024SCoexistence,
  author  = {Hao Xu and Chia-Min Chung and Mingpu Qin and Ulrich Schollw\"{o}ck and Steven R. White and Shiwei Zhang},
  journal = {Science},
  title   = {{Coexistence of superconductivity with partially filled stripes in the Hubbard model}},
  year    = {2024},
  number  = {6696},
  pages   = {eadh7691},
  volume  = {384},
  doi     = {10.1126/science.adh7691},
  ranking = {rank5},
  url     = {https://www.science.org/doi/abs/10.1126/science.adh7691},
}

@Article{Sinha2022PRBFinite,
  author    = {Aritra Sinha and Marek M. Rams and Piotr Czarnik and Jacek Dziarmaga},
  journal   = {Phys. Rev. B},
  title     = {{Finite-temperature tensor network study of the Hubbard model on an infinite square lattice}},
  year      = {2022},
  month     = {nov},
  number    = {19},
  pages     = {195105},
  volume    = {106},
  doi       = {10.1103/physrevb.106.195105},
  fjournal  = {Physical Review B},
  publisher = {American Physical Society ({APS})},
}

@Article{Jiang2020PRRGround,
  author    = {Jiang, Yi-Fan and Zaanen, Jan and Devereaux, Thomas P. and Jiang, Hong-Chen},
  journal   = {Phys. Rev. Res.},
  title     = {{Ground state phase diagram of the doped Hubbard model on the four-leg cylinder}},
  year      = {2020},
  month     = {Jul},
  pages     = {033073},
  volume    = {2},
  doi       = {10.1103/PhysRevResearch.2.033073},
  issue     = {3},
  numpages  = {14},
  publisher = {American Physical Society},
  url       = {https://link.aps.org/doi/10.1103/PhysRevResearch.2.033073},
}

@Article{Li2024PRLTime,
  author    = {Li, Jheng-Wei and Gleis, Andreas and von Delft, Jan},
  journal   = {Phys. Rev. Lett.},
  title     = {{Time-Dependent Variational Principle with Controlled Bond Expansion for Matrix Product States}},
  year      = {2024},
  month     = {Jul},
  pages     = {026401},
  volume    = {133},
  doi       = {10.1103/PhysRevLett.133.026401},
  issue     = {2},
  numpages  = {9},
  publisher = {American Physical Society},
  ranking   = {rank3},
  url       = {https://link.aps.org/doi/10.1103/PhysRevLett.133.026401},
}

@Article{Chen2017PRBSeries,
  author    = {Bin-Bin Chen and Yun-Jing Liu and Ziyu Chen and Wei Li},
  journal   = {Phys. Rev. B},
  title     = {Series-expansion thermal tensor network approach for quantum lattice models},
  year      = {2017},
  month     = {apr},
  number    = {16},
  pages     = {161104},
  volume    = {95},
  doi       = {10.1103/physrevb.95.161104},
  fjournal  = {Physical Review B},
  publisher = {American Physical Society ({APS})},
}

@Article{Dong2017PRBBilayer,
  author    = {Yong-Liang Dong and Lei Chen and Yun-Jing Liu and Wei Li},
  journal   = {Phys. Rev. B},
  title     = {Bilayer linearized tensor renormalization group approach for thermal tensor networks},
  year      = {2017},
  month     = {apr},
  number    = {14},
  pages     = {144428},
  volume    = {95},
  doi       = {10.1103/physrevb.95.144428},
  fjournal  = {Physical Review B},
  publisher = {American Physical Society ({APS})},
}

@Article{Keimer2015Nquantum,
  author    = {B. Keimer and S. A. Kivelson and M. R. Norman and S. Uchida and J. Zaanen},
  journal   = {Nature},
  title     = {From quantum matter to high-temperature superconductivity in copper oxides},
  year      = {2015},
  month     = {feb},
  number    = {7538},
  pages     = {179--186},
  volume    = {518},
  doi       = {10.1038/nature14165},
  publisher = {Springer Science and Business Media {LLC}},
  ranking   = {rank5},
}

@Article{Wietek2021PRXStripes,
  author    = {Alexander Wietek and Yuan-Yao He and Steven R. White and Antoine Georges and E. Miles Stoudenmire},
  journal   = {Phys. Rev. X},
  title     = {{Stripes, Antiferromagnetism, and the Pseudogap in the Doped Hubbard Model at Finite Temperature}},
  year      = {2021},
  month     = {jul},
  number    = {3},
  pages     = {031007},
  volume    = {11},
  doi       = {10.1103/physrevx.11.031007},
  fjournal  = {Physical Review X},
  publisher = {American Physical Society ({APS})},
  ranking   = {rank4},
}

@Article{Li2023PRLTangent,
  author    = {Qiaoyi Li and Yuan Gao and Yuan-Yao He and Yang Qi and Bin-Bin Chen and Wei Li},
  journal   = {Phys. Rev. Lett.},
  title     = {{Tangent Space Approach for Thermal Tensor Network Simulations of the 2D Hubbard Model}},
  year      = {2023},
  month     = {jun},
  number    = {22},
  pages     = {226502},
  volume    = {130},
  doi       = {10.1103/physrevlett.130.226502},
  fjournal  = {Physical Review Letters},
  publisher = {American Physical Society ({APS})},
}

@Article{Qu2024PRLPhase,
  author    = {Qu, Dai-Wei and Li, Qiaoyi and Gong, Shou-Shu and Qi, Yang and Li, Wei and Su, Gang},
  journal   = {Phys. Rev. Lett.},
  title     = {{Phase Diagram, $d$-Wave Superconductivity, and Pseudogap of the $t\ensuremath{-}{t}^{\ensuremath{'}}\ensuremath{-}J$ Model at Finite Temperature}},
  year      = {2024},
  month     = {Dec},
  pages     = {256003},
  volume    = {133},
  doi       = {10.1103/PhysRevLett.133.256003},
  issue     = {25},
  numpages  = {8},
  publisher = {American Physical Society},
  url       = {https://link.aps.org/doi/10.1103/PhysRevLett.133.256003},
}

@Article{Wang2023PRLPlaquette,
  author    = {Wang, Junsen and Li, Han and Xi, Ning and Gao, Yuan and Yan, Qing-Bo and Li, Wei and Su, Gang},
  journal   = {Phys. Rev. Lett.},
  title     = {{Plaquette Singlet Transition, Magnetic Barocaloric Effect, and Spin Supersolidity in the Shastry-Sutherland Model}},
  year      = {2023},
  month     = {Sep},
  pages     = {116702},
  volume    = {131},
  doi       = {10.1103/PhysRevLett.131.116702},
  issue     = {11},
  numpages  = {7},
  publisher = {American Physical Society},
  url       = {https://link.aps.org/doi/10.1103/PhysRevLett.131.116702},
}

@Article{Gleis2023PRLControlled,
  author    = {Andreas Gleis and Jheng-Wei Li and Jan von Delft},
  journal   = {Phys. Rev. Lett.},
  title     = {{Controlled Bond Expansion for Density Matrix Renormalization Group Ground State Search at Single-Site Costs}},
  year      = {2023},
  month     = {jun},
  number    = {24},
  pages     = {246402},
  volume    = {130},
  doi       = {10.1103/physrevlett.130.246402},
  publisher = {American Physical Society ({APS})},
}

@Article{Mai2023NCRobust,
  author    = {Mai, Peizhi and Nichols, Nathan S. and Karakuzu, Seher and Bao, Feng and Del Maestro, Adrian and Maier, Thomas A. and Johnston, Steven},
  journal   = {Nature Communications},
  title     = {{Robust charge-density-wave correlations in the electron-doped single-band Hubbard model}},
  year      = {2023},
  issn      = {2041-1723},
  month     = may,
  number    = {1},
  volume    = {14},
  doi       = {10.1038/s41467-023-38566-7},
  publisher = {Springer Science and Business Media LLC},
  ranking   = {rank4},
}

@Article{Qu2024PRLBilayer,
  author    = {Qu, Xing-Zhou and Qu, Dai-Wei and Chen, Jialin and Wu, Congjun and Yang, Fan and Li, Wei and Su, Gang},
  journal   = {Phys. Rev. Lett.},
  title     = {{Bilayer ${t\text{\ensuremath{-}}J\text{\ensuremath{-}}J}_{\ensuremath{\perp}}$ Model and Magnetically Mediated Pairing in the Pressurized Nickelate ${\mathrm{La}}_{3}{\mathrm{Ni}}_{2}{\mathrm{O}}_{7}$}},
  year      = {2024},
  month     = {Jan},
  pages     = {036502},
  volume    = {132},
  doi       = {10.1103/PhysRevLett.132.036502},
  issue     = {3},
  numpages  = {6},
  publisher = {American Physical Society},
  url       = {https://link.aps.org/doi/10.1103/PhysRevLett.132.036502},
}

@Article{Choi1972CJoMPositive,
  author    = {Choi, Man-Duen},
  journal   = {Canadian Journal of Mathematics},
  title     = {{Positive Linear Maps on C*-Algebras}},
  year      = {1972},
  number    = {3},
  pages     = {520-529},
  volume    = {24},
  doi       = {10.4153/CJM-1972-044-5},
  publisher = {Cambridge University Press},
}

@Article{Verstraete2004PRLMatrix,
  author    = {Verstraete, F. and Garc\'{\i}a-Ripoll, J. J. and Cirac, J. I.},
  journal   = {Phys. Rev. Lett.},
  title     = {{Matrix Product Density Operators: Simulation of Finite-Temperature and Dissipative Systems}},
  year      = {2004},
  month     = {Nov},
  pages     = {207204},
  volume    = {93},
  doi       = {10.1103/PhysRevLett.93.207204},
  issue     = {20},
  numpages  = {4},
  publisher = {American Physical Society},
  url       = {https://link.aps.org/doi/10.1103/PhysRevLett.93.207204},
}

@Article{Vidal2003PRLEfficient,
  author    = {Vidal, Guifr\'e},
  journal   = {Phys. Rev. Lett.},
  title     = {{Efficient Classical Simulation of Slightly Entangled Quantum Computations}},
  year      = {2003},
  month     = {Oct},
  pages     = {147902},
  volume    = {91},
  doi       = {10.1103/PhysRevLett.91.147902},
  issue     = {14},
  numpages  = {4},
  publisher = {American Physical Society},
  url       = {https://link.aps.org/doi/10.1103/PhysRevLett.91.147902},
}

@Article{Vidal2004PRLEfficient,
  author    = {Vidal, Guifr\'e},
  journal   = {Phys. Rev. Lett.},
  title     = {{Efficient Simulation of One-Dimensional Quantum Many-Body Systems}},
  year      = {2004},
  month     = {Jul},
  pages     = {040502},
  volume    = {93},
  doi       = {10.1103/PhysRevLett.93.040502},
  issue     = {4},
  numpages  = {4},
  publisher = {American Physical Society},
  url       = {https://link.aps.org/doi/10.1103/PhysRevLett.93.040502},
}

@Article{Lu2024PRLThermodynamic,
  author    = {Lu, Hongyu and Chen, Bin-Bin and Wu, Han-Qing and Sun, Kai and Meng, Zi Yang},
  journal   = {Phys. Rev. Lett.},
  title     = {{Thermodynamic Response and Neutral Excitations in Integer and Fractional Quantum Anomalous Hall States Emerging from Correlated Flat Bands}},
  year      = {2024},
  month     = {Jun},
  pages     = {236502},
  volume    = {132},
  doi       = {10.1103/PhysRevLett.132.236502},
  issue     = {23},
  numpages  = {7},
  publisher = {American Physical Society},
  url       = {https://link.aps.org/doi/10.1103/PhysRevLett.132.236502},
}

@Article{Gleis2022PRBProjector,
  author    = {Gleis, Andreas and Li, Jheng-Wei and von Delft, Jan},
  journal   = {Phys. Rev. B},
  title     = {Projector formalism for kept and discarded spaces of matrix product states},
  year      = {2022},
  month     = {Nov},
  pages     = {195138},
  volume    = {106},
  doi       = {10.1103/PhysRevB.106.195138},
  issue     = {19},
  numpages  = {14},
  publisher = {American Physical Society},
  url       = {https://link.aps.org/doi/10.1103/PhysRevB.106.195138},
}

@Article{Hubbard1963PRSAElectron,
  author  = {Hubbard, J.},
  journal = {Proc. R. Soc. A},
  title   = {Electron {Correlations} in {Narrow Energy Bands}},
  year    = {1963},
  month   = nov,
  number  = {1365},
  pages   = {238--257},
  volume  = {276},
  doi     = {10.1098/rspa.1963.0204},
}

@Article{Arovas2022ARoCMPHubbard,
  author    = {Arovas, Daniel P. and Berg, Erez and Kivelson, Steven A. and Raghu, Srinivas},
  journal   = {Annual Review of Condensed Matter Physics},
  title     = {{The Hubbard Model}},
  year      = {2022},
  issn      = {1947-5462},
  month     = mar,
  number    = {1},
  pages     = {239--274},
  volume    = {13},
  doi       = {10.1146/annurev-conmatphys-031620-102024},
  publisher = {Annual Reviews},
}

@Article{White1992PRLDensity,
  author    = {White, Steven R.},
  journal   = {Phys. Rev. Lett.},
  title     = {{Density Matrix Formulation for Quantum Renormalization Groups}},
  year      = {1992},
  month     = {Nov},
  pages     = {2863--2866},
  volume    = {69},
  doi       = {10.1103/PhysRevLett.69.2863},
  issue     = {19},
  numpages  = {0},
  publisher = {American Physical Society},
  url       = {https://link.aps.org/doi/10.1103/PhysRevLett.69.2863},
}

@Article{Schollwoeck2005RMPdensity,
  author    = {Schollw\"ock, U.},
  journal   = {Rev. Mod. Phys.},
  title     = {The density-matrix renormalization group},
  year      = {2005},
  month     = {Apr},
  pages     = {259--315},
  volume    = {77},
  doi       = {10.1103/RevModPhys.77.259},
  issue     = {1},
  numpages  = {0},
  publisher = {American Physical Society},
  url       = {https://link.aps.org/doi/10.1103/RevModPhys.77.259},
}

@Article{Schollwoeck2011APDensity,
  author  = {Ulrich {Schollw{\"o}ck}},
  journal = {Ann. Phys.},
  title   = {{The Density-Matrix Renormalization Group in the Age of Matrix Product States}},
  year    = {2011},
  issn    = {0003-4916},
  number  = {1},
  pages   = {96-192},
  volume  = {326},
  doi     = {https://doi.org/10.1016/j.aop.2010.09.012},
  url     = {https://www.sciencedirect.com/science/article/pii/S0003491610001752},
}

@Article{Blankenbecler1981PRDMonte,
  author    = {Blankenbecler, R. and Scalapino, D. J. and Sugar, R. L.},
  journal   = {Phys. Rev. D},
  title     = {{Monte Carlo Calculations of Coupled Boson-Fermion Systems. I}},
  year      = {1981},
  month     = {Oct},
  pages     = {2278--2286},
  volume    = {24},
  doi       = {10.1103/PhysRevD.24.2278},
  issue     = {8},
  numpages  = {0},
  publisher = {American Physical Society},
  url       = {https://link.aps.org/doi/10.1103/PhysRevD.24.2278},
}

@Article{Scalapino1981PRBMonte,
  author    = {Scalapino, D. J. and Sugar, R. L.},
  journal   = {Phys. Rev. B},
  title     = {{Monte Carlo calculations of coupled boson-fermion systems. II}},
  year      = {1981},
  month     = {Oct},
  pages     = {4295--4308},
  volume    = {24},
  doi       = {10.1103/PhysRevB.24.4295},
  issue     = {8},
  numpages  = {0},
  publisher = {American Physical Society},
  url       = {https://link.aps.org/doi/10.1103/PhysRevB.24.4295},
}

@Article{Dagotto1994RMPCorrelated,
  author    = {Dagotto, Elbio},
  journal   = {Rev. Mod. Phys.},
  title     = {Correlated electrons in high-temperature superconductors},
  year      = {1994},
  month     = {Jul},
  pages     = {763--840},
  volume    = {66},
  doi       = {10.1103/RevModPhys.66.763},
  issue     = {3},
  numpages  = {0},
  publisher = {American Physical Society},
  ranking   = {rank5},
  url       = {https://link.aps.org/doi/10.1103/RevModPhys.66.763},
}

@Article{Gao2024PRBDouble,
  author    = {Gao, Yuan and Zhang, Chuandi and Xiang, Junsen and Yu, Dehong and Lu, Xingye and Sun, Peijie and Jin, Wentao and Su, Gang and Li, Wei},
  journal   = {Phys. Rev. B},
  title     = {Double magnon-roton excitations in the triangular-lattice spin supersolid},
  year      = {2024},
  month     = {Dec},
  pages     = {214408},
  volume    = {110},
  doi       = {10.1103/PhysRevB.110.214408},
  issue     = {21},
  numpages  = {8},
  publisher = {American Physical Society},
  url       = {https://link.aps.org/doi/10.1103/PhysRevB.110.214408},
}

@Article{Iglovikov2015PRBGeometry,
  author    = {Iglovikov, V. I. and Khatami, E. and Scalettar, R. T.},
  journal   = {Phys. Rev. B},
  title     = {{Geometry dependence of the sign problem in quantum Monte Carlo simulations}},
  year      = {2015},
  month     = {Jul},
  pages     = {045110},
  volume    = {92},
  doi       = {10.1103/PhysRevB.92.045110},
  issue     = {4},
  numpages  = {13},
  publisher = {American Physical Society},
  url       = {https://link.aps.org/doi/10.1103/PhysRevB.92.045110},
}

@Misc{Devos2025TensorKit.jl,
  author        = {Lukas Devos and Jutho Haegeman},
  title         = {{TensorKit.jl: A Julia package for large-scale tensor computations, with a hint of category theory}},
  year          = {2025},
  archiveprefix = {arXiv},
  eprint        = {2508.10076},
  primaryclass  = {cs.MS},
  url           = {https://arxiv.org/abs/2508.10076},
}

@Article{Qin2022ARoCMPHubbard,
  author    = {Qin, Mingpu and Sch\"{a}fer, Thomas and Andergassen, Sabine and Corboz, Philippe and Gull, Emanuel},
  journal   = {Annual Review of Condensed Matter Physics},
  title     = {{The Hubbard Model: A Computational Perspective}},
  year      = {2022},
  issn      = {1947-5462},
  month     = mar,
  number    = {1},
  pages     = {275--302},
  volume    = {13},
  doi       = {10.1146/annurev-conmatphys-090921-033948},
  publisher = {Annual Reviews},
}

@Article{Li2011PRLLinearized,
  author    = {Li, Wei and Ran, Shi-Ju and Gong, Shou-Shu and Zhao, Yang and Xi, Bin and Ye, Fei and Su, Gang},
  journal   = {Phys. Rev. Lett.},
  title     = {{Linearized Tensor Renormalization Group Algorithm for the Calculation of Thermodynamic Properties of Quantum Lattice Models}},
  year      = {2011},
  month     = {Mar},
  pages     = {127202},
  volume    = {106},
  doi       = {10.1103/PhysRevLett.106.127202},
  issue     = {12},
  numpages  = {4},
  publisher = {American Physical Society},
  url       = {https://link.aps.org/doi/10.1103/PhysRevLett.106.127202},
}

@Article{Chen2026SBFractional,
  author    = {Chen, Jialin and Li, Qiaoyi and Wang, Xiaoyu and Li, Wei},
  journal   = {Science Bulletin},
  title     = {{Fractional Chern insulator and quantum anomalous Hall crystal in twisted MoTe2}},
  year      = {2026},
  issn      = {2095-9273},
  month     = jan,
  doi       = {10.1016/j.scib.2026.01.014},
  publisher = {Elsevier BV},
}

@Article{Speidel2006JoCTaCAutomatic,
  author    = {Speidel, Joshua A. and Banfelder, Jason R. and Mezei, Mihaly},
  journal   = {Journal of Chemical Theory and Computation},
  title     = {{Automatic Control of Solvent Density in Grand Canonical Ensemble Monte Carlo Simulations}},
  year      = {2006},
  issn      = {1549-9626},
  month     = jul,
  number    = {5},
  pages     = {1429--1434},
  volume    = {2},
  doi       = {10.1021/ct0600363},
  publisher = {American Chemical Society (ACS)},
}

@Article{Lv2025NCQuantum,
  author    = {Lv, Enze and Xi, Ning and Jin, Yuliang and Li, Wei},
  journal   = {Nature Communications},
  title     = {{Quantum supercritical regime with universal magnetocaloric scaling in Ising magnets}},
  year      = {2025},
  issn      = {2041-1723},
  month     = nov,
  number    = {1},
  volume    = {16},
  doi       = {10.1038/s41467-025-65651-w},
  publisher = {Springer Science and Business Media LLC},
}

@Article{Orus2014AoPpractical,
  author    = {Or\'us, Rom\'an},
  journal   = {Annals of Physics},
  title     = {A practical introduction to tensor networks: Matrix product states and projected entangled pair states},
  year      = {2014},
  issn      = {0003-4916},
  month     = oct,
  pages     = {117--158},
  volume    = {349},
  doi       = {10.1016/j.aop.2014.06.013},
  publisher = {Elsevier BV},
}

@Article{Mak2022NNSemiconductor,
  author    = {Mak, Kin Fai and Shan, Jie},
  journal   = {Nature Nanotechnology},
  title     = {Semiconductor moir\'e materials},
  year      = {2022},
  issn      = {1748-3395},
  month     = jul,
  number    = {7},
  pages     = {686--695},
  volume    = {17},
  doi       = {10.1038/s41565-022-01165-6},
  publisher = {Springer Science and Business Media LLC},
}

@Article{Andrei2020NMGraphene,
  author    = {Andrei, Eva Y. and MacDonald, Allan H.},
  journal   = {Nature Materials},
  title     = {Graphene bilayers with a twist},
  year      = {2020},
  issn      = {1476-4660},
  month     = nov,
  number    = {12},
  pages     = {1265--1275},
  volume    = {19},
  doi       = {10.1038/s41563-020-00840-0},
  publisher = {Springer Science and Business Media LLC},
}

@Misc{Coleman2015Heavy,
  author        = {Piers Coleman},
  title         = {{Heavy Fermions and the Kondo Lattice: a 21st Century Perspective}},
  year          = {2015},
  archiveprefix = {arXiv},
  eprint        = {1509.05769},
  primaryclass  = {cond-mat.str-el},
  url           = {https://arxiv.org/abs/1509.05769},
}

@Article{Si2010SHeavy,
  author    = {Si, Qimiao and Steglich, Frank},
  journal   = {Science},
  title     = {{Heavy Fermions and Quantum Phase Transitions}},
  year      = {2010},
  issn      = {1095-9203},
  month     = sep,
  number    = {5996},
  pages     = {1161--1166},
  volume    = {329},
  doi       = {10.1126/science.1191195},
  publisher = {American Association for the Advancement of Science (AAAS)},
}

@Article{Yang2016RoPiPTwo,
  author    = {Yang, Yi-feng},
  journal   = {Reports on Progress in Physics},
  title     = {Two-fluid model for heavy electron physics},
  year      = {2016},
  issn      = {1361-6633},
  month     = may,
  number    = {7},
  pages     = {074501},
  volume    = {79},
  doi       = {10.1088/0034-4885/79/7/074501},
  publisher = {IOP Publishing},
}

@Book{Horn2012Matrix,
  author    = {Roger A. Horn and Charles R. Johnson},
  publisher = {Cambridge University Press},
  title     = {Matrix Analysis},
  year      = {2012},
  address   = {Cambridge},
  edition   = {2nd},
  isbn      = {978-0521548236},
}

@Book{Lee2012GTiMIntroduction,
  author    = {Lee, John M.},
  publisher = {Springer New York},
  title     = {Introduction to Smooth Manifolds},
  year      = {2012},
  isbn      = {9781441999825},
  doi       = {10.1007/978-1-4419-9982-5},
  issn      = {0072-5285},
  journal   = {Graduate Texts in Mathematics},
}

@Article{Miles2022PREDynamical,
  author    = {Miles, Cole and Cohen-Stead, Benjamin and Bradley, Owen and Johnston, Steven and Scalettar, Richard and Barros, Kipton},
  journal   = {Phys. Rev. E},
  title     = {{Dynamical tuning of the chemical potential to achieve a target particle number in grand canonical Monte Carlo simulations}},
  year      = {2022},
  month     = {Apr},
  pages     = {045311},
  volume    = {105},
  doi       = {10.1103/PhysRevE.105.045311},
  issue     = {4},
  numpages  = {9},
  publisher = {American Physical Society},
  url       = {https://link.aps.org/doi/10.1103/PhysRevE.105.045311},
}

@Article{CohenStead2024SPCSmoQyDQMC.jl,
  author    = {Benjamin Cohen-Stead and Sohan Malkaruge Costa and James Neuhaus and Andy Tanjaroon Ly and Yutan Zhang and Richard Scalettar and Kipton Barros and Steven Johnston},
  journal   = {SciPost Phys. Codebases},
  title     = {{SmoQyDQMC.jl: A flexible implementation of determinant quantum Monte Carlo for Hubbard and electron-phonon interactions}},
  year      = {2024},
  pages     = {29},
  doi       = {10.21468/SciPostPhysCodeb.29},
  publisher = {SciPost},
  url       = {https://scipost.org/10.21468/SciPostPhysCodeb.29},
}

@Misc{Zhang2025Finite,
  author        = {Yintai Zhang and Aritra Sinha and Marek M. Rams and Jacek Dziarmaga},
  title         = {{Finite temperature dopant-induced spin reorganization explored via tensor networks in the two-dimensional $t$-$J$ model}},
  year          = {2025},
  archiveprefix = {arXiv},
  eprint        = {2510.04756},
  primaryclass  = {cond-mat.str-el},
  url           = {https://arxiv.org/abs/2510.04756},
}

@Article{Xu2023PRXObservation,
  author    = {Xu, Fan and Sun, Zheng and Jia, Tongtong and Liu, Chang and Xu, Cheng and Li, Chushan and Gu, Yu and Watanabe, Kenji and Taniguchi, Takashi and Tong, Bingbing and Jia, Jinfeng and Shi, Zhiwen and Jiang, Shengwei and Zhang, Yang and Liu, Xiaoxue and Li, Tingxin},
  journal   = {Phys. Rev. X},
  title     = {{Observation of Integer and Fractional Quantum Anomalous {Hall} Effects in Twisted Bilayer ${\mathrm{MoTe}}_{2}$}},
  year      = {2023},
  month     = {Sep},
  pages     = {031037},
  volume    = {13},
  doi       = {10.1103/PhysRevX.13.031037},
  issue     = {3},
  numpages  = {12},
  publisher = {American Physical Society},
  url       = {https://link.aps.org/doi/10.1103/PhysRevX.13.031037},
}

@Article{Park2023NObservation,
  author    = {Park, Heonjoon and Cai, Jiaqi and Anderson, Eric and Zhang, Yinong and Zhu, Jiayi and Liu, Xiaoyu and Wang, Chong and Holtzmann, William and Hu, Chaowei and Liu, Zhaoyu and Taniguchi, Takashi and Watanabe, Kenji and Chu, Jiun-Haw and Cao, Ting and Fu, Liang and Yao, Wang and Chang, Cui-Zu and Cobden, David and Xiao, Di and Xu, Xiaodong},
  journal   = {Nature},
  title     = {Observation of fractionally quantized anomalous {Hall} effect},
  year      = {2023},
  issn      = {1476-4687},
  month     = aug,
  number    = {7981},
  pages     = {74--79},
  volume    = {622},
  doi       = {10.1038/s41586-023-06536-0},
  publisher = {Springer Science and Business Media LLC},
}

@Article{Cai2023NSignatures,
  author    = {Cai, Jiaqi and Anderson, Eric and Wang, Chong and Zhang, Xiaowei and Liu, Xiaoyu and Holtzmann, William and Zhang, Yinong and Fan, Fengren and Taniguchi, Takashi and Watanabe, Kenji and Ran, Ying and Cao, Ting and Fu, Liang and Xiao, Di and Yao, Wang and Xu, Xiaodong},
  journal   = {Nature},
  title     = {{Signatures of fractional quantum anomalous Hall states in twisted MoTe2}},
  year      = {2023},
  issn      = {1476-4687},
  month     = jun,
  number    = {7981},
  pages     = {63--68},
  volume    = {622},
  doi       = {10.1038/s41586-023-06289-w},
  publisher = {Springer Science and Business Media LLC},
}

@Article{Czarnik2012PRBProjected,
  author     = {Czarnik, P. and Cincio, L. and Dziarmaga, J.},
  journal    = {Phys. Rev. B},
  title      = {{Projected Entangled Pair States at Finite Temperature: Imaginary Time Evolution with Ancillas}},
  year       = {2012},
  month      = {Dec},
  pages      = {245101},
  volume     = {86},
  doi        = {10.1103/PhysRevB.86.245101},
  issue      = {24},
  numpages   = {6},
  publisher  = {American Physical Society},
  url        = {https://link.aps.org/doi/10.1103/PhysRevB.86.245101},
}

@Article{Czarnik2016PRBVariational,
  author    = {Piotr Czarnik and Marek M. Rams and Jacek Dziarmaga},
  journal   = {Phys. Rev. B},
  title     = {{Variational tensor network renormalization in imaginary time: Benchmark results in the Hubbard model at finite temperature}},
  year      = {2016},
  month     = {dec},
  number    = {23},
  pages     = {235142},
  volume    = {94},
  doi       = {10.1103/physrevb.94.235142},
  fjournal  = {Physical Review B},
  publisher = {American Physical Society ({APS})},
}

\end{document}